\documentclass[a4paper, 12pt]{article}
\usepackage{latexsym, amsmath, amsfonts, amssymb}
\usepackage{tikz}
\usetikzlibrary{decorations.pathmorphing, decorations.markings}
\usepackage[american]{babel}
\usepackage{graphicx}
\usepackage{bbm}
\usepackage{cite}
\usepackage{enumitem}
\usepackage{mathrsfs}
\usepackage[colorlinks=true, citecolor=blue!90!black, linkcolor=blue!90!black, linktocpage=true, urlcolor=red!70!black]{hyperref}

\renewcommand{\baselinestretch}{1.2}
\newcommand{\Vol}{\mathrm{Vol}}

\newcommand{\ds}{\displaystyle}
\newcommand{\wt}{\widetilde}

\newcommand{\matht}[1]{{\ensuremath{\boldsymbol{#1}}}}
\newcommand{\tps}[2]{\texorpdfstring{#1}{#2}}
\newcommand{\subf}[2]{{\small\begin{tabular}[t]{@{}c@{}}#1\\#2\end{tabular}}}

\newcommand{\eg}{\textit{e.g.}}

\newcommand{\ie}{\textit{i.e.}}

\numberwithin{equation}{section}

\newcommand{\nn}{\nonumber}

\newcommand{\smat}[1]{\big( \begin{smallmatrix} #1 \end{smallmatrix} \big)}
\newcommand{\be}{\begin{equation}} \newcommand{\ee}{\end{equation}}
\newcommand{\bea}{\begin{equation} \begin{aligned}} \newcommand{\eea}{\end{aligned} \end{equation}}

\newcommand{\cA}{\mathcal{A}}

\newcommand{\cD}{\mathcal{D}}

\newcommand{\cG}{\mathcal{G}}
\newcommand{\cH}{\mathcal{H}}

\newcommand{\cL}{\mathcal{L}}
\newcommand{\cM}{\mathcal{M}}

\newcommand{\cO}{\mathcal{O}}

\newcommand{\cS}{\mathcal{S}}
\newcommand{\cT}{\mathcal{T}}

\newcommand{\bC}{\mathbb{C}}

\newcommand{\bH}{\mathbb{H}}

\newcommand{\bR}{\mathbb{R}}

\newcommand{\bZ}{\mathbb{Z}}

\newcommand{\sT}{\mathsf{T}}

\newcommand{\unit}{\mathbbm{1}}

\DeclareMathOperator{\Tr}{Tr}

\DeclareMathOperator{\re}{\mathbb{R}e}
\DeclareMathOperator{\im}{\mathbb{I}m}

\begin{document}
\thispagestyle{empty}
\fontsize{12pt}{16pt}
\begin{flushright}
	SISSA  14/2024/FISI
\end{flushright}
\vspace{13mm}  
\begin{center}
	{\huge  Entanglement asymmetry \\[.6em] in conformal field theory and holography}
	\\[13mm]
    	{\large Francesco Benini$^{\, a,b,c}$, \, Victor Godet$^{\, a,b}$, Amartya Harsh Singh$^{\, a,b}$}
	
	\bigskip
	{\it
		$^a$ SISSA, Via Bonomea 265, 34136 Trieste, Italy \\[-0.0em]
		$^b$ INFN, Sezione di Trieste, Via Valerio 2, 34127 Trieste, Italy \\[.2em]
		$^c$ ICTP, Strada Costiera 11, 34151 Trieste, Italy
	}
\end{center}

\bigskip

\begin{abstract}
\noindent
Entanglement asymmetry is a measure of symmetry breaking in quantum subsystems, inspired by quantum information theory, particularly suited to study out-of-equilibrium states. We study the entanglement asymmetry of a class of excited ``coherent states'' in conformal quantum field theories with a $U(1)$ symmetry, employing Euclidean path-integral methods with topological symmetry defects and the replica formalism. We compute, at leading order in perturbation theory, the asymmetry for a variety of subsystems, including finite spherical subregions in flat space, in finite volume, and at positive temperature. We also study its Lorentzian time evolution, showcasing the dynamical restoration of the symmetry due to thermalization, as well as the presence of a quantum Mpemba effect. Our results are universal, and apply in any number of dimensions. We also show that the perturbative entanglement asymmetry is related to the Fisher information metric, which has a known holographic dual called Hollands--Wald canonical energy, and that it is captured by the AdS bulk charge contained in the entanglement wedge.
\end{abstract}

\newpage
\pagenumbering{arabic}
\setcounter{page}{1}
\setcounter{footnote}{0}
\renewcommand{\thefootnote}{\arabic{footnote}}

{\renewcommand{\baselinestretch}{.88} \parskip=0pt
\setcounter{tocdepth}{2}
\tableofcontents}


\section{Introduction}
\label{sec: intro}

Symmetry and symmetry breaking are key concepts in quantum field theory (QFT). Indeed, often it is only through the lenses of symmetry breaking that we decode the complicated dynamics of many-body systems or strongly-coupled QFTs. The traditional language to describe symmetry breaking is the Landau paradigm, \ie, the analysis of the expectation values of local order parameters. On the other hand, quantum information theory has been playing over the years an increasingly important role in many-body systems and quantum field theory. For instance, in the context of QFT, quantum information can be used to derive monotonicity and positivity results \cite{Wall:2011hj, Casini:2012ei, Faulkner:2016mzt}. In the context of AdS/CFT, quantum information allows us to understand the emergence of the bulk from the CFT \cite{Faulkner:2013ica, Almheiri:2014lwa}, to obtain the Page curve for toy models of evaporating black holes \cite{Penington:2019npb, Almheiri:2019psf}, and it appears to be well-suited to study dynamics far from equilibrium \cite{deBoer:2023lrd}.

Recently a new quantum information observable --- dubbed \emph{entanglement asymmetry} --- has been introduced in \cite{Ares:2022koq} to detect and quantify symmetry breaking in quantum subsystems, in particular out of equilibrium. It is a modification of entanglement entropy. Given a reduced density matrix $\rho$ that describes a pure or mixed state as seen from within a subsystem $A$, and given a $U(1)$ action on the Hilbert space, \cite{Ares:2022koq} proposes to compare $\rho$ with the average of $\rho$ over the adjoint action of $U(1)$, that we call $\rho_Q$.%
\footnote{In fact this works for any group $G$: discrete, Abelian, and non-Abelian.}
Precisely, the entanglement asymmetry $\Delta S$ is defined as the difference between the entanglement entropies of $\rho$ and $\rho_Q$, and it turns out to be equivalent to a relative entropy:
\be
\Delta S = \Tr \bigl( \rho \log \rho \bigr) - \Tr \bigl( \rho_Q \log \rho_Q \bigr) = S \bigl( \rho \,\Vert\, \rho_Q \bigr) \,.
\ee
This is a good measure of symmetry breaking: by known properties of relative entropy, $\Delta S$ is non-negative, and it vanishes if and only if the symmetry is unbroken in $A$.

Entanglement asymmetry was used in \cite{Ares:2022koq} to investigate a quantum version of the Mpemba effect \cite{Mpemba:1969, Lu:2017, Kumar:2020} in spin chains. This is the phenomenon that excited symmetry-breaking states can relax and restore the symmetry faster than states in which the symmetry is less broken.%
\footnote{Aristotle observed, some 2,300 years ago, that ``to cool hot water quickly, begin by putting it in the sun''. The quantum Mpemba effect has been observed experimentally in ion traps \cite{Joshi:2024sup}.}
Such a counter-intuitive behavior is only possible far from equilibrium. Subsequent works that studied entanglement asymmetry and the Mpemba effect include \cite{Ares:2023kcz, Bertini:2023ysg, Ferro:2023sbn, Capizzi:2023yka, Capizzi:2023xaf, Rylands:2023yzx, Murciano:2023qrv, Chen:2023gql, Caceffo:2024jbc, Fossati:2024xtn, Yamashika:2024hpr, Liu:2024kzv, Chalas:2024wjz, Ares:2024nkh, Turkeshi:2024juo, Klobas:2024png}. 

In this paper we are interested in computing entanglement asymmetry in conformal quantum field theories (CFTs) with a $U(1)$ symmetry in general dimensions $d$, with an eye towards holography. In QFT, entanglement entropy is known to suffer from UV divergences. For entanglement asymmetry instead, being a relative entropy, the divergences cancel out and we expect to obtain a well-defined observable. In CFTs the vacuum does not spontaneously break the symmetry, thus we ought to consider some excited state. Here we study a class of ``coherent states'', partially motivated by holography \cite{Botta-Cantcheff:2015sav, Christodoulou:2016nej, Faulkner:2017tkh}:
\be
|\Psi\rangle = e^{i \lambda V} |0\rangle \,,
\ee
obtained by acting on the vacuum with $e^{i\lambda V}$ in Euclidean time, where $V$ is a primary local operator of charge 1.%
\footnote{Similar states have been studied in \cite{Chen:2023gql}.}
We choose to work in a perturbative expansion in $\lambda$, and show that $\Delta S = \lambda^2 \Delta S^{(2)} + O(\lambda^4)$. We compute the leading contribution $\Delta S^{(2)}$ in three different ways:
\begin{enumerate}[noitemsep,topsep=0pt]
\item By first computing the Renyi asymmetries
\be
\Delta S_n = \frac1{1-n} \Bigl( \log \Tr \rho_Q^n - \log \Tr \rho^n \Bigr)
\ee
using the replica method of \cite{Holzhey:1994we, Calabrese:2004eu, Calabrese:2009qy}, and then taking the limit $n \to 1$.

\item By exploiting an integral representation for the leading perturbative contribution to relative entropy, also called Fisher information \cite{Lashkari:2015hha, Faulkner:2017tkh}.

\item By computing the Renyi asymmetries in terms of correlation functions of $V$ and $V^\dag$ that include twist operators \cite{Calabrese:2004eu}.
\end{enumerate}
The result is a ``universal expression'' for the entanglement asymmetry, that depends on the dimension $\Delta$ of $V$ and on the location of its insertion in Euclidean space.

It turns out that both $\Delta S^{(2)}$ and $\Delta S^{(2)}_n$ behave as two-point functions. This allows us (for instance as in \cite{Cardy:2016fqc}) to exploit conformal transformations to determine the entanglement asymmetry in a variety of geometries, for different choices of the subsystem $A$: for the ``Rindler'' geometry in which $A$ is an infinite half-space; for finite spherical subregions $A$ in infinite volume as well as in finite volume on $S^{d-1}$; for spherical subregions on the hyperbolic plane $\bH^{d-1}$ at finite temperature; for finite intervals on the real line at finite temperature in $d=2$.

Following \cite{Calabrese:2006rx, Cardy:2016fqc}, we are also able to use conformal transformations and analytic continuation to determine the time evolution of entanglement asymmetry, starting from a coherent state (that we regard as originating from a local quench). As expected from thermalization, the asymmetry of a subsystem decreases with time since the excited state relaxes. Interestingly, however, we observe that a universal quantum Mpemba effect takes place. It would be interesting to identify an underlying mechanism responsible for the Mpemba effect in CFTs (as done in \cite{Rylands:2023yzx} for integrable systems).

In the context of AdS/CFT, what is the holographic dual to entanglement asymmetry? At leading order in $\lambda$, entanglement asymmetry is Fisher information, and the latter has a known holographic dual \cite{Lashkari:2015hha} given by a suitable integral of the so-called symplectic flux in the entanglement wedge. We observe that such a quantity can be recast in a suggestive form:
\be
\Delta S^{(2)} = 2\pi \, \partial_{\tau_0} \biggl( \, \int_{\wt A} \star\, F - \int_A \star\, F\biggr) \,.
\ee
Here $F$ is the field strength of the bulk gauge field dual to the $U(1)$ symmetry of the CFT, $A$ is the chosen spatial subsystem at the boundary, $\wt A$ is the Ryu--Takayanagi surface (homologous to $A$) that describes entanglement entropy, while $\tau_0$ is the boundary location of the insertion of $V$ along the modular flow. This formula thus contains all the ingredients that we expect to be important in the holographic evaluation of entanglement asymmetry. We hope to be able to address the higher-order terms, and to obtain a better physical understanding of this formula, in future work.


\section{Entanglement asymmetry}
\label{sec: review asymmetry}

Let us review the definition of entanglement asymmetry proposed in \cite{Ares:2022koq}.%
\footnote{The authors of \cite{Ma:2021zgf} considered a very similar quantity, but for global states.}
Consider a quantum system which can be divided into two parts $A$ and $B$ in such a way that the Hilbert space factorizes: $\cH = \cH_A \otimes \cH_B$. Given a (normalized) density matrix $\rho_\text{tot}$, which could represent a pure ($\rho_\text{tot} = | \psi \rangle \langle \psi|$) or a mixed state, one constructs the reduced density matrix on $A$, namely $\rho = \Tr_B \rho_\text{tot}$, and its entanglement entropy is given by the von~Neumann entropy $S = - \Tr( \rho \log \rho)$. Consider the case that there exists a charge operator $Q$ that generates a $U(1)$ action on the Hilbert space,%
\footnote{In most cases, and in this paper too, the $U(1)$ action is a global symmetry of the quantum system. However the Hamiltonian of the system does not enter into the definition of entanglement asymmetry and therefore one could consider $U(1)$ actions that are not symmetries, as for instance in \cite{Fossati:2024xtn}.}
and that the charge operator can be factorized as well:
\be
Q = Q_A \otimes \unit_B + \unit_A \otimes Q_B \,.
\ee
If the state $\rho_\text{tot}$ is an eigenstate of $Q$, namely $[\rho_\text{tot}, Q]=0$ and thus the symmetry is unbroken, then $[\rho, Q_A]=0$ and thus $\rho$ takes a block-diagonal form in a basis of eigenspaces of $Q_A$. Then the entanglement entropy $S$ can be resolved into the contributions $S(q)$ from each charge sector: this is called symmetry-resolved entanglement entropy \cite{Laflorencie:2014cem, Goldstein:2017bua, Xavier:2018kqb}. On the other hand, if the symmetry is broken in the subsystem $A$ and thus $[\rho, Q_A] \neq 0$, one can quantify the amount of breaking in the following way. One constructs a projected reduced density matrix
\be
\label{def projected rho 1}
\rho_Q = \sum\nolimits_{q \in \bZ} \Pi_q \, \rho \, \Pi_q \,,
\ee
where $\Pi_q$ is the projector to the eigenspace of $Q_A$ with charge $q$. This is essentially the matrix $\rho$ with all off-diagonal blocks set to zero. The quantity (\ref{def projected rho 1}) can be recast in a more elegant form by using the Fourier transform of the projector $\Pi_q = \int_0^{2\pi} \! \frac{d\alpha}{2\pi} \, e^{i\alpha(Q_A - q)}$, hence
\be
\label{def projected rho 2}
\rho_Q = \int_0^{2\pi} \frac{d\alpha}{2\pi} e^{-i\alpha Q_A} \, \rho \, e^{i \alpha Q_A} \,.
\ee
This is the average of $\rho$ over the adjoint action of the group $U(1)$, and it naturally generalizes to discrete as well as non-Abelian groups \cite{Ferro:2023sbn, Capizzi:2023yka, Capizzi:2023xaf, Fossati:2024xtn}. Notice that $\rho_Q$ is a density matrix, namely it is Hermitian, semi-positive definite, and $\Tr \rho_Q = 1$. The entanglement asymmetry is defined as the difference between the entanglement entropies of $\rho_Q$ and $\rho$ \cite{Ares:2022koq}:
\be
\Delta S = S[\rho_Q] - S[\rho] = \Tr (\rho \log \rho) - \Tr (\rho_Q \log \rho_Q) \,.
\ee
One easily verifies that the entanglement asymmetry is equivalent to the relative entropy of $\rho$ with respect to $\rho_Q$ \cite{Ma:2021zgf}:
\be
\Delta S = \Tr \bigl( \rho \, (\log \rho - \log \rho_Q) \bigr) = S( \rho \,\Vert\, \rho_Q) \,.
\ee
This makes $\Delta S$ a good measure of symmetry breaking. Indeed, as it follows from the properties of relative entropy: $\Delta S \geq 0$, and $\Delta S = 0$ if and only if $\rho = \rho_Q$ namely if $[\rho, Q_A] = 0$ so that the symmetry is unbroken in $A$.

The entanglement asymmetry can be obtained from Renyi entanglement asymmetries using the replica trick \cite{Holzhey:1994we, Calabrese:2004eu}. The Renyi entanglement asymmetries are defined as
\be
\label{def renyi asymmetry}
\Delta S_n = \frac1{1-n} \log \frac{ \Tr ( \rho_Q^n) }{ \Tr ( \rho^n) } \,.
\ee
The entanglement asymmetry is obtained from the limit $\Delta S = \lim_{n\to 1} \Delta S_n$. The moments $\Tr (\rho_Q^n)$ are easily obtained from (\ref{def projected rho 2}). By a change of integration variables, they can be written as \cite{Ma:2021zgf, Ares:2022koq}:
\be
\Tr (\rho_Q^n) = \int_0^{2\pi} \frac{d\gamma_1 \ldots d\gamma_{n-1}}{(2\pi)^{n-1}} \, X_n(A, \gamma)
\ee
where we defined the so-called charged moments
\be
X_n(A, \gamma) = \Tr \Bigl[ \rho \, e^{i\gamma_1 Q_A} \, \rho \, e^{i\gamma_2 Q_A} \,\cdots\, \rho \, e^{i \gamma_{n-1} Q_A} \, \rho \, e^{-i (\gamma_1 + \ldots + \gamma_{n-1}) Q_A } \Bigr] \,.
\ee
We could formally define $\gamma_n \equiv - (\gamma_1 + \ldots + \gamma_{n-1})$ so that $\sum_{j=1}^n \gamma_j = 0$. Thus, the charged moments implement all insertions of elements of $U(1)$ that multiply to the identity.

We are interested in studying entanglement asymmetry in quantum field theories, following \cite{Calabrese:2004eu}. In particular, we will use the path integral over replicas in order to define and compute traces of density matrices. It is well known that the definition of entanglement entropy we presented is plagued by UV divergences when used in quantum field theory, as opposed to quantum mechanics. The situation improves for entanglement asymmetry (as it does in general for relative entropies) because the divergences cancel among the numerator and denominator of (\ref{def renyi asymmetry}), at least as long as the symmetry breaking is spontaneous, or explicit but soft.

We consider states $\rho_\text{tot}$ that have a Euclidean path-integral representation. In particular, given a spatial manifold $\cM$, the state $\rho_\text{tot}$ is produced by the path integral on some Euclidean manifold, possibly with operator insertions, with a cut along a copy of $\cM$. We call this geometry $\cG$. Schematically, the matrix elements of such a state are
\be
\langle \phi_- | \rho_\text{tot} | \phi_+ \rangle = \frac1Z \int_{\substack{ \text{geometry $\cG$} \\ \varphi(0^-, \cM) = \phi_- \\ \varphi(0^+, \cM) = \phi_+ }} \cD\varphi\, (\dots) \, e^{-S[\varphi]} \,.
\ee
Here $(t_\text{E}, \cM)$ are Euclidean coordinates with $t_\text{E}$ the Euclidean time and the cut at $t_\text{E} = 0$, $\phi_\mp$ are Dirichlet boundary conditions for the fields, $(\dots)$ are possible operator insertions, while $Z$ is the partition function on $\cG$ with the cut closed so as to guarantee that $\Tr \rho_\text{tot} = 1$. The reduced density matrix $\rho$ is obtained by partially closing the cut along $B$ while keeping it open along $A \subset \cM$. The moments of $\rho$ are computed by
\be
\Tr(\rho^n) = \frac{Z_n(A)}{Z^n} \,.
\ee
Here $Z_n(A)$ is the Euclidean path integral of the theory on a manifold $\cG_n$ obtained by gluing together $n$ identical copies of $\cG$ (possibly with operator insertions) along the cut $A$, in such a way that the lower part of the cut on the $n$-th copy is glued to the upper part of the cut on the $(n+1)$-th copy. Then $Z \equiv Z_1(A)$ is the path integral on $\cG$ (possibly with operator insertions) with the cut completely closed.

In order to compute the charged moments $X_n(A,\gamma)$ of $\rho_Q$ we need to insert the operators $e^{i\gamma_j Q_A}$ into the trace \cite{Fossati:2024xtn}. In the path-integral description they are the topological codimension-one surfaces $U_\gamma[A]$, also called symmetry defects, that implement the $U(1)$ action on the Hilbert space \cite{Gaiotto:2014kfa}, placed along the cut $A$. They are
\be
U_\gamma[A] = \exp\biggl( i \gamma \int_A \star \, J \biggr) \qquad\text{labelled by } \gamma \in [0, 2\pi) \cong U(1) \,,
\ee
where $J$ is the conserved $U(1)$ current operator, while $\star$ is the Hodge star operation. Thus
\be
X_n(A, \gamma) = \frac{Z_n(A, \gamma) }{ Z^n } \,.
\ee
Here $Z_n(A,\gamma)$ is the path integral on $\cG_n$ with the insertion of topological symmetry defects $U_{\gamma_j}[A]$ along each of the gluings from one replica to the next. In the original geometry $\cG$ it may appear that the operator $U_\alpha[A]$ has a boundary along $\partial A$, and in general the boundaries of symmetry defect operators are \emph{not} topological. However on the covering geometry $\cG_n$ there are $n$ symmetry defects that join along $\partial A$ and with parameters $\gamma_j$ that sum up to $0 \in U(1)$, therefore in this case the junction of defects is topological as well.%
\footnote{In the presence of an 't~Hooft anomaly for the $U(1)$ symmetry, one may be worried that the path-integral definition of Renyi asymmetries is plagued by phase ambiguities: indeed one needs to resolve the junction of $n$ symmetry defects into a network of trivalent junctions, and different choices are related by nontrivial phases in the presence of an 't~Hooft anomaly (we thank Giovanni Galati for raising this issue). In two dimensions each connected component of $A$ is an interval with two ends. We insist that the resolution of the junction be specular on the two ends: in this way, if one changes the resolution one obtains conjugate phases from the two ends and they cancel each other. In higher dimensions each connected component of $A$ has a connected boundary, and we similarly insist that the resolution be ``constant'' along that boundary. We leave a more detailed study of possible effects of 't~Hooft anomalies and higher-group structures on entanglement asymmetry for the future.}
Notice that $Z_1(A,\gamma) = Z$.

Eventually, the path-integral formula for the Renyi asymmetries is
\be
\label{Renyi asymmetry QFT}
\Delta S_n = \frac1{1-n} \log \frac{ \frac1{(2\pi)^{n-1}} \int\! d\vec\gamma \; Z_n(A, \gamma) }{ Z_n(A) } \,,
\ee
where $d\vec \gamma$ is a shorthand notation for $d\gamma_1 \dots d\gamma_{n-1}$. Similar partition functions with topological defects were recently considered in \cite{Gutperle:2024rvo} to study entanglement entropy in symmetric product orbifold CFTs.

There exists an alternative computational approach in which one considers $n$ replicas of the theory, as opposed to $n$ replicas of the Euclidean geometry. In this approach one computes the path integral of the theory in the presence of twist fields placed along $\partial A$ that implement the gluing of one copy of the theory to the next as one crosses $A$ \cite{Calabrese:2004eu}. In the context of entanglement asymmetry, the twist fields are dressed by the endpoints of a suitable symmetry defect placed along $A$ \cite{Capizzi:2023yka, Fossati:2024xtn}. We illustrate and utilize this approach in Appendix~\ref{app: asymmetry from twist}.


\section{Asymmetry for CFTs in the Rindler geometry}
\label{sec: Rindler}

We are interested in the computation of entanglement asymmetry in CFTs with a $U(1)$ symmetry.%
\footnote{Entanglement asymmetry in CFTs in which the symmetry is explicitly broken has been studied in \cite{Fossati:2024xtn}.}
Since the vacuum of a CFT does not exhibit spontaneous symmetry breaking, we should consider some excited state. In this paper we study a class of ``coherent states'' obtained by acting on the vacuum with a local operator $\cO(z) = e^{i \lambda V(z)}$ inserted along Euclidean time.%
\footnote{More precisely, one should define $\cO$ as a smeared operator $\cO = e^{i \!\int\! d^dz\, \lambda(z) \, V(z)}$ obtained by integrating $V(z)$ on a region of spacetime in the Euclidean past. The resulting excited coherent state is $|\psi\rangle = e^{i \!\int\! d^dz\, \lambda(z) \, V(z)} |0\rangle$.
This construction parallels the holographic setup discussed in Sec.~\ref{sec: holography}. As the smearing function $\lambda(z)$ is concentrated on a smaller and smaller support, limiting to a Dirac delta function, the operator $\cO$ develops divergences. There are no divergences, however, at linear order in $\lambda$ and therefore, with some abuse, we will keep using the notation $\cO(z) = e^{i\lambda V(z)}$ for $\cO$, with the above understanding in mind.}
Here $V$ is a primary operator with fixed charge under $Q$, that for simplicity we take to be $1$, while $\lambda$ is a small real parameter. The state is thus
\be
\label{def coherent states}
|\psi\rangle = \cO(z_-) \, |0\rangle = e^{i \lambda V(z_-)} \, |0\rangle \;,\qquad\quad
\rho_\text{tot} = |\psi \rangle \langle \psi| = e^{i \lambda V(z_-)} \, |0 \rangle \langle 0 | \, e^{-i \lambda V^\dag(z_+)} \,.
\ee
Here $z_-$ is the insertion point in the Euclidean past, while $z_+$ is its specular image under the inversion of Euclidean time. We give a graphical representation of the path integral in Fig.~\ref{fig: path-integral preparation}. Since the operator $\cO$ does not have definite charge, the state breaks the symmetry. This class of states is particularly natural from the point of view of holography, as we discuss in Section~\ref{sec: holography}, and has already been studied for instance in \cite{Botta-Cantcheff:2015sav, Christodoulou:2016nej, Faulkner:2017tkh}.

\begin{figure}[t]
\centering
\begin{tikzpicture}
	\node at (-4, 0) {$\langle \phi_- | \, \rho \, | \phi_+ \rangle \quad = \quad \dfrac1Z$};
	\draw [->] (-1.8, 0) -- (1.8, 0);
	\draw [->] (0, -1.2) -- (0, 1.2);
	\draw [red!80!black, thick, double] (-1, 0) -- (1, 0);
	\filldraw (-1, 0) circle [radius = 0.05]; \filldraw (1, 0) circle [radius = 0.05];
	\node at (-0.4, 0.32) {\small$\phi_+$};
	\node at (-0.4, -0.3) {\small$\phi_-$};
	\filldraw [blue] (0.6, -0.7) circle [radius = 0.04] node[shift={(-0.15, -0.25)}, black] {\small$z_-$};
	\node[right] at (0.65, -0.55) {\small$e^{i\lambda V}$};
	\draw [blue, thick] (0.6, 0.7) circle [radius = 0.04] node[shift={(-0.15, -0.25)}, black] {\small$z_+$};
	\node[right] at (0.65, 0.85) {\small$e^{-i\lambda V^\dag}$};
\end{tikzpicture}
\caption{\label{fig: path-integral preparation}%
Path-integral preparation of the reduced density matrix $\rho$ for the excited coherent states $|\psi\rangle$ studied in this paper.}
\end{figure}

Given a spatial manifold $\cM$, the state is prepared by the Euclidean path integral on $\bR \times \cM$ (where the first factor is Euclidean time $t_\text{E}$) with a cut at $t_\text{E} = 0$, the insertion of $\cO(z_-)$ at a point $z_-$ with $t_\text{E} < 0$, and the insertion of $\cO^\dag(z_+)$ where $z_+$ is the point obtained from $z_-$ by $t_\text{E} \mapsto - t_\text{E}$. We can then write the Renyi asymmetry (\ref{Renyi asymmetry QFT}) in terms of correlation functions:
\be
\label{Renyi asymmetry as correlator}
\Delta S_n = \frac1{1-n} \log \frac{ \frac1{(2\pi)^{n-1}} \int\! d\vec\gamma \; \bigl\langle \cO_1 \cO_1^\dag \cL_{\gamma_1} \cdots \cO_n \cO_n^\dag \cL_{\gamma_n} \bigr\rangle_{\cG_n} }{ \bigl\langle \cO_1 \cO_1^\dag \cdots \cO_n \cO^\dag_n \bigr\rangle_{\cG_n} } \,.
\ee
Here $\cO_j \equiv \cO(z_{j-})$ is the operator inserted in the $j$-th replica, while $\cL_{\gamma_j}$ is the topological symmetry defects placed along the gluing from the $j$-th to the $(j+1)$-th replica. See also \cite{Chen:2023gql} where some Renyi asymmetries were discussed as correlation functions.

In this paper we will content ourselves with computing the entanglement asymmetry at leading order in a perturbative expansion at small $\lambda$, \ie, in a limit in which the symmetry is only slightly broken. Calling $\Delta$ the dimension of the primary $V$, notice that $\lambda$ has dimensions of length$^\Delta$ and thus we study a limit in which $\lambda$ is small compared to the Euclidean distance between the insertion point $z_-$ and the cut $A$. In the denominator of (\ref{Renyi asymmetry as correlator}) we find
\be
\bigl\langle \cO_1 \cO_1^\dag \ldots \cO_n \cO_n^\dag \bigr\rangle_{\cG_n} = 1 + \lambda^2 \sum_{j,k} \bigl\langle V(z_{j-}) \, V^\dag(z_{k+}) \bigr\rangle_{\cG_n} + O(\lambda^4) \,.
\ee
Notice that there are no terms with odd powers of $\lambda$ because of charge conservation in the vacuum. Besides, at order $\lambda^2$ we never pick up the product of two $V$'s or two $V^\dag$'s at the same point and therefore we do not have to worry about divergences. In the numerator of (\ref{Renyi asymmetry as correlator}) we have a correlator in the presence of the symmetry defects $\cL_{\gamma_j}$. Since they are topological, we can swipe them through the replicas until they are all moved to the same sheet, and then we can collapse them on top of each other. Since the parameters sum up to zero, they disappear. However, when they cross a local operator, the correlator picks up a phase. We thus find
\be
\bigl\langle \cO_1 \cO_1^\dag \cL_{\gamma_1} \,\ldots\, \cO_n \cO_n^\dag \cL_{\gamma_n} \bigr\rangle_{\cG_n} = 1 + \lambda^2 \sum_{j,k} e^{i \theta_{jk}} \, \bigl\langle V(z_{j-}) \, V^\dag(z_{k+}) \bigr\rangle_{\cG_n} + O(\lambda^4) \,,
\ee
where $\theta_{jk} = \sum_{i=j}^{k-1} \gamma_i$ if $k>j$, or $\theta_{jk} = - \sum_{i=k}^{j-1} \gamma_i$ if $k<j$, or $\theta_{jk} = 0$ if $j=k$. Once we integrate in $d\vec \gamma$, all terms in the summation with a nontrivial phase are projected out, therefore the summation reduces to $j=k$:
\be
\frac1{(2\pi)^{n-1}} \!\int\! d\vec\gamma \; \bigl\langle \cO_1 \cO_1^\dag \cL_{\gamma_1} \,\dots\, \cO_n \cO_n^\dag \cL_{\gamma_n} \bigr\rangle_{\cG_n} = 1 + \lambda^2 \sum_j \bigl\langle V(z_{j-}) V^\dag(z_{j+}) \bigr\rangle_{\cG_n} + O(\lambda^4) \,.
\ee
In other words, the integral projects to covering geometries in which the total charge vanishes copy by copy, not just globally in the union of copies. From (\ref{Renyi asymmetry as correlator}) we eventually obtain
\be
\label{Renyi asymmetry 2-point func formula}
\Delta S_n = \frac{\lambda^2}{n-1} \sum_{j \neq k}^n \bigl\langle V(z_{j-}) V^\dag(z_{k+}) \bigr\rangle_{\cG_n} + O(\lambda^4) \,.
\ee
This gives the leading perturbative contribution to the Renyi asymmetries.

\subsection{2d CFTs and replicas}
\label{sec: replica}

\begin{figure}[t]
\centering
\begin{tikzpicture}
	\draw [black!60!white, ->] (-2, 0) -- (2.5, 0);
	\draw [black!60!white, ->] (0, -1.5) -- (0, 1.5);
	\draw [red!80!black, thick, double] (0,0) -- (2.3, 0) node [above, pos = 0.7, black] {\small$A$};
	\draw [dashed, black!70!white] (0,0) -- (1.2, -1) node [black, pos = 0.35, shift={(-0.15, -0.15)}] {\small$r$};
	\filldraw (0,0) circle [radius = 0.05];
	\draw [->] (-176: 1) arc [radius = 1, start angle = -176, end angle = -45] node [below left, black, pos = 0.5] {\small$\tau_0$};
	\filldraw [blue] (1.2, -1) circle [radius = 0.04] node [black, right] {\small$z_-$} node [black, shift={(-110: 0.35)}] {\small$\cO$};
	\draw [blue, thick] (1.2, 1) circle [radius = 0.04] node [black, right] {\small$z_+$} node [black, shift = {(110: 0.35)}] {\small$\cO^\dag$};
	\node at (2.3, 1.6) {\footnotesize$z$};
	\draw (2.1, 1.8) -- (2.1, 1.4) -- (2.5, 1.4);
	\draw [->] (3.8, 0.8) arc [start angle = 120, end angle = 60, radius = 1.7] node [pos = 0.5, above] {\footnotesize$\zeta = z^{1/n}$};
\begin{scope}[shift={(9,0)}]
	\draw (-2, 0) -- (2, 0);
	\draw (0, -1.5) -- (0, 1.5);
	\draw [thick, red!80!black, dashed] (0, 0.05) -- (2, 0.05) node [right, black] {\small$\cL_n$};
	\draw [thick, red!80!black, dashed] (-0.05, 0) -- (-0.05, 1.5) node [above, black] {\small$\cL_1$};
	\draw [thick, red!80!black, dashed] (0, -0.05) -- (-2, -0.05) node [left, black] {\small$\cL_2$};
	\draw [thick, red!80!black, dashed] (0.05, 0) -- (0.05, -1.5) node [below, black] {\small$\cL_{n-1}$};
	\draw [blue, thick] (1, 0.5) circle [radius = 0.04] node [black, right] {\small$\zeta_{1+}$};
	\filldraw [blue] (0.5, 1) circle [radius = 0.04] node [black, above] {\small$\zeta_{1-}$};
	\draw [blue, thick] (-0.5, 1) circle [radius = 0.04] node [black, above] {\small$\zeta_{2+}$};
	\filldraw [blue] (-1, 0.5) circle [radius = 0.04] node [black, left] {\small$\zeta_{2-}$};
	\draw [thick, dotted] (-1, -0.5) arc [radius = 1.12, start angle = -153.4, end angle = -116.6];
	\draw [blue, thick] (0.5, -1) circle [radius = 0.04] node [black, below] {\small$\zeta_{n+}$};
	\filldraw [blue] (1, -0.5) circle [radius = 0.04] node [black, right] {\small$\zeta_{n-}$};
	\node at (2.3, 1.6) {\footnotesize$\zeta$};
	\draw (2.1, 1.8) -- (2.1, 1.4) -- (2.5, 1.4);
\end{scope}
\end{tikzpicture}
\caption{\label{fig: z and zeta planes}%
Left: physical $z$-plane of the Rindler geometry, with insertions of $\cO$ at $z_- = r\, e^{-i(\pi - \tau_0)}$ and $\cO^\dag$ at $z_+ = z_-^*$. Right: covering $\zeta$-plane, in the case $n=4$. The conformal map is $\zeta = z^{1/n}$.}
\end{figure}
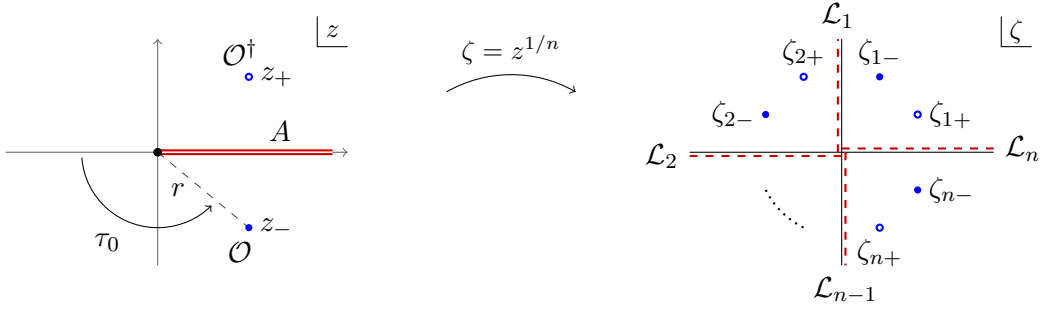

Two-point functions on the replicated geometries $\cG_n$ are non-trivial because the branching loci $\partial A$ create curvature singularities. The two-dimensional case is special because, in certain cases, the singularities can be removed by a conformal transformation. We thus consider a 2d CFT, and the case that $A$ is the semi-infinite line to the right of the origin. Using a complex coordinate $z = z_2 + i z_1$ (where $z_1$ is Euclidean time and $z_2$ is space) we take $A = \{ z \in \bR_+\}$. We call this case the \emph{Rindler geometry}, as in Fig.~\ref{fig: z and zeta planes} left (we will consider other geometries in Section~\ref{sec:conformalmaps}). We can then compute correlators on $\cG_n$ by exploiting a conformal covering map $\zeta = z^{1/n}$. Two-point functions transform as
\be
\bigl\langle V(z_{j-}) \, V^\dag(z_{k+}) \bigr\rangle_{\cG_n} = \biggl\lvert \frac{d\zeta}{dz}(\zeta_{j-}) \, \frac{d\zeta}{dz}(\zeta_{k+}) \biggr\rvert^\Delta \, \bigl\langle V(\zeta_{j-}) V^\dag(\zeta_{k+}) \bigr\rangle_\bC \;,
\ee
where $\Delta$ is the dimension of $V$. We parametrize the insertion points $z_\pm$ as
\be
\label{labelling of insertions in Rindler}
z_- = r \, e^{-i(\pi - \tau_0)} \;,\qquad z_+ = r\, e^{i(\pi-\tau_0)} \;,\qquad r>0 \;,\quad 0 < \tau_0 < \pi \,.
\ee
On the covering $\zeta$-plane we take
\be
\zeta_{j-} = r^{1/n} e^{i \left[ 2\pi\left( j - \frac12 \right) + \tau_0 \right]/n} \;,\qquad\qquad \zeta_{j+} = r^{1/n} e^{i \left[ 2\pi \left( j - \frac12 \right) - \tau_0 \right] /n} \;.
\ee
The twist lines $\cL_j$ (before being removed) run along the semi-infinite lines with $\arg(\zeta) = 2\pi j/n$, as depicted in Fig.~\ref{fig: z and zeta planes} right. The two-point functions on the covering geometries $\cG_n$ are
\be
\bigl\langle V(z_{j-}) \, V^\dag(z_{k+}) \bigr\rangle_{\cG_n} = \frac1{n^{2\Delta} \; r^{2\Delta(n-1)/n} \, \bigl\lvert \zeta_{j-} - \zeta_{k+} \bigr\rvert^{2\Delta} } \,.
\ee
Writing the Renyi asymmetry as $\Delta S_n = \lambda^2 \Delta S_n^{(2)} + O(\lambda^4)$, the leading term from (\ref{Renyi asymmetry 2-point func formula}) is
\be
\label{Renyi asymmetries 2d CFT}
\Delta S_n^{(2)} = \frac{n^{-2\Delta}}{n-1} \, \frac1{r^{2\Delta}} \sum_{j\neq k}^n \frac1{ \bigl\lvert 2 \sin \bigl( \frac{\tau_0 + (j - k)\pi }n \bigr) \bigr\rvert^{2\Delta} } \,.
\ee

In order to compute the leading contribution $\Delta S^{(2)}$ to the entanglement asymmetry we need to analytically continue (\ref{Renyi asymmetries 2d CFT}) to complex values of $n$ and then take the limit $n \to 1$. This means that we need to perform the sum explicitly. We were able to do that only for some integer values of $2\Delta$, using the Mellin transform
\be
\label{Mellin transform}
\frac1{ \sin(\pi x)} = \frac1\pi \int_\bR ds \, \frac{e^{sx}}{1+e^s} \qquad\qquad\text{for } 0 < \re(x) < 1 \,.
\ee
In Appendix~\ref{app: asymmetry from replicas} we perform the computation for $\Delta = \frac12, 1, \frac32, 2$. For instance, for $\Delta =1$ the Renyi asymmetry of the Rindler geometry takes the form
\be
\label{Renyi asymm Delta = 1}
\Delta = 1: \qquad\qquad \Delta S^{(2)}_n = \frac1{4r^2} \, \frac{n}{n-1} \biggl( \frac1{ \sin^2(\tau_0)} - \frac1{ n^2 \sin^2\bigl( \frac{\tau_0}n \bigr)} \biggr) \,,
\ee
therefore the entanglement asymmetry is
\be
\Delta = 1: \qquad\qquad \Delta S^{(2)}_\text{Rindler} = \frac1{2 r^2 \sin^2(\tau_0)} \biggl( 1 - \frac{\tau_0}{ \tan(\tau_0) } \biggr) \,. \hspace{0.9cm}
\ee
In the next section we will determine $\Delta S^{(2)}_\text{Rindler}$ for all $\Delta \in \bR_+$.

\subsection{Asymmetry as relative entropy}
\label{sec: rel entropy}

We can make progress in the computation of the leading contribution to the entanglement asymmetry of coherent states in a perturbative expansion in $\lambda$ --- in general dimension $d$ --- by using that entanglement asymmetry is a relative entropy.

Given two density matrices $\rho_0$, $\rho_1$, their relative entropy is defined as
\be
S(\rho_1 \,\Vert\, \rho_0) = \Tr (\rho_1 \log \rho_1) - \Tr (\rho_1 \log \rho_0) \,.
\ee
Consider a continuous family $\rho(\lambda) = \rho_0 + \lambda\, \delta\rho + O(\lambda^2)$, normalized so that $\Tr \rho(\lambda) = 1$. Then, the relative entropy $S\bigl( \rho(\lambda) \,\Vert\, \rho_0 \bigr)$ starts at second order in $\lambda$ and at that order it is given by
\be
\label{formula quadratic relative entropy}
\frac12 \, \frac{d^2}{d\lambda^2} \, S\bigl( \rho(\lambda) \,\Vert\, \rho_0 \bigr) = F\bigl( \delta\rho, \delta\rho)_{\rho_0} \,,
\ee
where the quantity $F(\delta\rho_1, \delta\rho_2)_{\rho_0}$ is symmetric in its arguments and is called the Fisher information metric around $\rho_0$. There exists an integral formula for it, see \eg{} Appendix B of \cite{Faulkner:2017tkh}:
\be
\label{formula Fisher information}
F\bigl( \delta\rho_1, \delta\rho_2)_{\rho_0} = \frac14 \int_{-\infty}^\infty \frac{ds}{1+\cosh(s)} \Tr \Bigl[ \delta \rho_1 \; \rho_0^{-\frac12 - \frac{is}{2\pi}} \; \delta\rho_2 \; \rho_0^{-\frac12 + \frac{is}{2\pi}} \Bigr] \,.
\ee
This formula is valid even if $\rho(\lambda)$ is not Hermitian, as long as it is normalized.

In our setup the reduced density matrix $\rho$ is
\be
\label{state rho}
\rho = \sigma + \lambda \, \sigma \, (iV - i V^\dag) + O(\lambda^2) \,,
\ee
where $\sigma = \Tr_B |0 \rangle \langle 0|$ is the reduced density matrix of the vacuum. Since the vacuum does not break the symmetry while $\sum_q \Pi_q \, \sigma V \, \Pi_q = 0$, we obtain that the projection $\rho_Q$ differs from the vacuum only at second order in $\lambda$: $\rho_Q = \sigma + O(\lambda^2)$. We already noticed that the entanglement asymmetry can be written as a relative entropy $\Delta S = S(\rho \,\Vert\, \rho_Q)$. Using again that $\sigma$ commutes with $Q_A$, we can also write the asymmetry as $\Delta S = S(\rho \,\Vert\, \sigma) - S(\rho_Q \,\Vert\, \sigma)$. Since $\rho_Q$ is equal to $\sigma$ at first order in $\lambda$, the second term does not contribute at leading order. We thus obtain
\be
\Delta S = \lambda^2 \, S^{(2)}(\rho \,\Vert\, \sigma) + O(\lambda^4) \,.
\ee

We apply this formula to the Rindler geometry in $d \geq 2$ dimensions. The spatial slice is $\bR^{d-1}$, the subsystem $A$ is a half-space, and the entangling surface $\partial A$ is a $(d-2)$-dimensional plane. We use a complex coordinate $z$ for the spatial direction orthogonal to the entangling surface and the Euclidean time direction, and $y_i$ for the other coordinates (if any). The state is obtained from the vacuum with the insertion of $\cO = e^{i\lambda V}$ at $z_-$ and $\cO^\dag = e^{-i \lambda V^\dag}$ at $z_+ = z_-^*$, both on the plane $y_i = 0$. We could more generally consider insertions at two generic points, not necessarily related by complex conjugation.%
\footnote{In this case the Euclidean path integral prepares a non-Hermitian matrix.}
We will be interested in the case
\be
\label{more general z_+-}
z_- = r \, e^{-i(\pi - \tau_-)} = r \, e^{i\theta_-} \;,\qquad\qquad z_+ = r \, e^{i(\pi - \tau_+)} = r \, e^{i \theta_+} \;.
\ee
The reduced density matrix $\sigma$ of the vacuum in the Rindler geometry is special because its modular Hamiltonian $K$ defined by $\sigma = e^{-K}$ is local and geometric: it generates a counterclockwise rotation of the complex coordinate $z$ around the entangling surface. We can thus write the state $\rho$ as
\bea
\rho &= e^{- \left( 1 - \frac{\theta_-}{2\pi} \right) K} \, e^{i\lambda V} \, e^{- \frac{\theta_- - \theta_+}{2\pi} K} \, e^{- i \lambda V^\dag} \, e^{- \frac{\theta_+}{2\pi} K} \\[-.2em]
&= \sigma + i \lambda \, \sigma \, e^{\frac{\theta_-}{2\pi} K} \, V \, e^{- \frac{\theta_-}{2\pi} K} - i \lambda \, \sigma \, e^{\frac{\theta_+}{2\pi} K} \, V^\dag \, e^{- \frac{\theta_+}{2\pi} K} + O(\lambda^2) \\
&= \sigma + i \lambda \, \sigma \, V(z_-) - i \lambda \, \sigma \, V^\dag(z_+) + O(\lambda^2) \,,
\eea
as already written in (\ref{state rho}). To this state we apply (\ref{formula quadratic relative entropy})--(\ref{formula Fisher information}). Taking into account charge conservation, we obtain two terms which however are equal using the change of variable $s \to -s$ and the cyclicity of the trace:
\be
\Delta S^{(2)}_\text{Rindler} = \frac12 \int\! \frac{ds}{1+\cosh(s)} \Tr \Bigl[ \sigma \, V^\dag(z_+) \, \sigma^{- \frac12 - \frac{is}{2\pi}} \, \sigma \, V(z_-) \, \sigma^{-\frac12 + \frac{is}{2\pi}} \Bigr] \,.
\ee
Shifting the integration contour as $s \to s + \pi i (1 - \varepsilon)$ where $\varepsilon$ is an infinitesimal positive quantity that specifies a contour prescription, the integral is recast as
\be
\label{integral with trace}
\Delta S^{(2)}_\text{Rindler} = -\frac14 \int \frac{ds}{\sinh^2\bigl( \frac s2 - i \varepsilon \bigr) } \Tr \Bigl[ \sigma \, \sigma^{- \frac{is}{2\pi}} \, V(z_-) \, \sigma^{ \frac{is}{2\pi}} \, V^\dag(z_+) \Bigr] \,.
\ee
The trace contains operators time-ordered with respect to modular time and is thus a two-point function. With $s=0$, if we take two generic ordered points $z_j = e^{u_j + i \theta_j}$ we have
\begin{multline}
\Tr \Bigl[ \sigma \, V(z_1) \, V^\dag(z_2) \Bigr] = \bigl\langle V(z_1) \, V^\dag(z_2) \bigr\rangle = \frac1{|z_1 - z_2|^{2\Delta}} = {} \\
{} = \frac1{ \Bigl[ 2 \, e^{u_1+u_2} \bigl( \cosh(u_1 - u_2) - \cos(\theta_1 - \theta_2) \bigr) \Bigr]^\Delta } \equiv G_\Delta(u_1, \theta_1, u_2, \theta_2) \;.
\end{multline}
Since $V(z_-) = e^{\frac{\theta_-}{2\pi} K} \, V(r,0) \, e^{- \frac{\theta_-}{2\pi} K}$ we see that $\sigma^{-\frac{is}{2\pi}} \,V(z_-)\, \sigma^{\frac{is}{2\pi}}$ has the net effect of shifting $\theta_- \to \theta_- + is$. The trace in (\ref{integral with trace}) is then $G_\Delta(u, \theta_- + is, u, \theta_+) = G_\Delta(u, \pi + \tau_- + is, u, \pi - \tau_+)$ in terms of our parametrization.
We obtain the formula:
\be
\label{asymmetry integral}
\Delta S^{(2)}_\text{Rindler} = - \int_{-\infty}^{+\infty} \frac{ds}{4 \, r^{2\Delta} \sinh^2\bigl( \frac s2 - i \varepsilon \bigr) \, \Bigl[ - 4 \sinh^2\bigl( \frac s2 - i \tau_0 \bigr) \Bigr]^\Delta }
\ee
where $\tau_0 = \frac12(\tau_+ + \tau_-)$.

\begin{figure}[t]
\centering
\includegraphics[height=4.5cm]{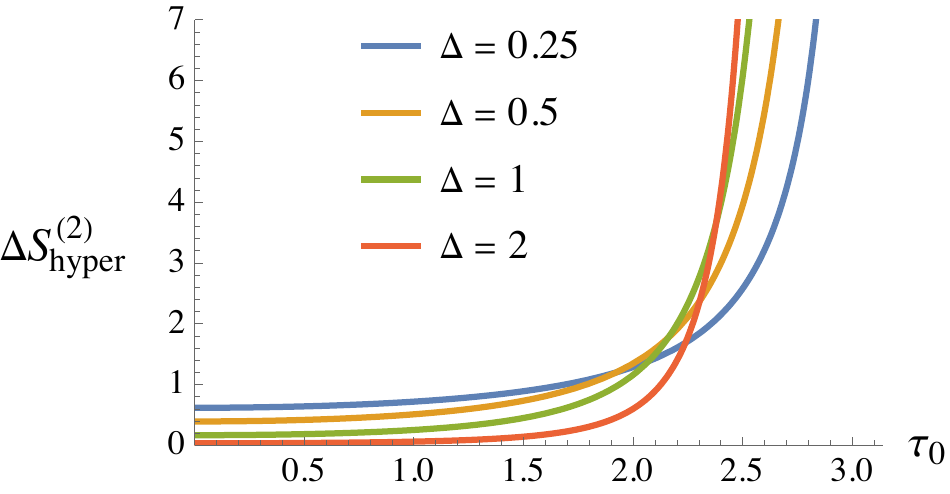}
\caption{\label{Fig:AsymmetryDifferentDelta}%
The function $I_\Delta \bigl( \tan\bigl( \frac{\tau_0}2 \bigr) \bigr)$ that describes $\Delta S^{(2)}_\text{Rindler}$ and $\Delta S^{(2)}_\text{hyper}$, plotted for different values of $\Delta$ as a function of $\tau_0 \in [0,\pi)$.}
\end{figure}

As we describe in Appendix~\ref{app:rindlerasymmetry}, the integral in (\ref{asymmetry integral}) can be analytically performed in terms of a hypergeometric function. Let us define the function:%
\footnote{Recall that $_2F_1(a,b,c;z) = \sum_{n=0}^\infty \frac{(a)_n (b)_n}{(c)_n} \, \frac{z^n}{n!}$ and in particular $_2F_1(a,b,c;0) = 1$. The analytic continuation for $|z| \geq 1$ has a branch cut from 1 to $\infty$ along the positive real axis.}
\be
\label{def function I}
I_\Delta(x) = \frac{\sqrt\pi \, \Gamma(\Delta+1) }{ 4^\Delta \, \Gamma\bigl( \Delta + \frac12 \bigr) } \, \biggl( 1 - \bigl(1-x^2 \bigr) \, \frac{\Delta}{ \Delta + \frac12} \; _2F_1\Bigl[ 1, \tfrac12 - \Delta, \tfrac32 + \Delta; - x^2 \Bigr] \biggr) \,.
\ee
Then
\be
\label{asymmetry Rindler general}
\Delta S^{(2)}_\text{Rindler} = \frac1{r^{2\Delta}} \; I_\Delta\Bigl( \tan \bigl( \tfrac{\tau_0}2 \bigr) \Bigr) \,.
\ee
The entanglement asymmetry of the Rindler geometry starts at a finite value for $\tau_0 = 0$ and increases monotonically with a divergence at $\tau_0 = \pi$ with the following behaviors:%
\footnote{They follow from $\ds \lim_{x \to 0} I_\Delta(x) = \frac{\sqrt\pi \, \Gamma(\Delta +1) }{ 2^{2\Delta +1} \, \Gamma(\Delta + 3/2) }$ and $\ds I_\Delta(x) \,\sim\, \frac{\pi \, \Delta}{ 16^\Delta} \, x^{2\Delta + 1}$ for $x \to +\infty$.}
\be
\lim_{\tau_0 \to 0} \, \Delta S^{(2)}_\text{Rindler} = \frac{ \sqrt{\pi} \, \Gamma(\Delta+1) }{ 2 \, (2r)^{2\Delta} \, \Gamma \bigl( \Delta + \frac32 \bigr) } \,,\qquad\qquad
\Delta S^{(2)}_\text{Rindler} \, \underset{\tau_0 \to \pi}{\sim} \, \frac{ 2\pi \Delta }{ (2r)^{2\Delta} \, \sin^{2\Delta+1}(\tau_0)} \,.
\ee
This is illustrated in Fig.~\ref{Fig:AsymmetryDifferentDelta} where we plot the function $I_\Delta\bigl( \tan(\tau_0/2) \bigr)$ for different values of $\Delta$. For semi-integer values of $\Delta$ we found alternative expressions for (\ref{asymmetry integral}) in terms of trigonometric functions. For $\Delta \in \frac12 + \bZ_{\geq 0}$ we found the expressions:
\be
\Delta S^{(2)}_\text{Rindler} = \frac{\pi \, \Delta}{(4r)^{2\Delta} \, \cos^{2\Delta+1}\bigl( \frac{\tau_0}2 \bigr) } \, \biggl[ 1 + \sum_{k=1}^{\Delta - \frac32} \frac{ \bigl( \Delta - \frac12 - k \bigr) \, \bigl( \Delta - \frac32 + k \bigr)! }{ k! \, \bigl( \Delta - \frac12 \bigr)! } \, \cos^{2k} \Bigl( \frac{\tau_0}2 \Bigr) \biggr] \;.
\ee
The summation is finite and contributes only for $\Delta \geq \frac52$.
For $\Delta \in \bZ_{\geq1}$ we found:
\be
\Delta S^{(2)}_\text{Rindler} = \frac{2\Delta}{ (2r)^{2\Delta} \, \sin^{2\Delta}(\tau_0)} \biggl[ 1 - \frac{ \tau_0}{\tan(\tau_0)} - \sum_{k=1}^{\Delta-1} \frac{(2k-2)!!}{(2k+1)!!} \, \sin^{2k}(\tau_0) \biggr] \;.
\ee
The summation is finite and contributes only for $\Delta \geq 2$. These expressions match with those we already found in $d=2$ for $\Delta = \frac12, 1, \frac32, 2$ using the replica method (see Appendix~\ref{app: asymmetry from replicas}).

\section{Other geometries, relaxation, and the Mpemba effect}
\label{sec:conformalmaps}

For the class of states (\ref{def coherent states}) and at leading order in $\lambda$, it is clear from (\ref{Renyi asymmetry 2-point func formula}) that both Renyi and entanglement asymmetries behave as two-point functions. This means that in a CFT we can exploit conformal transformations to determine the asymmetry in geometries other than the Rindler one.%
\footnote{This is the same as what was done in \cite{Calabrese:2004eu, Calabrese:2009qy, Cardy:2016fqc} for the entanglement entropy.}
In this section we will compute the entanglement asymmetry of finite spherical regions $A$ in arbitrary dimensions, both in infinite volume, in a finite spherical volume, and in the hyperbolic plane at finite temperature.%
\footnote{In $d=2$ dimensions, the hyperbolic plane at finite temperature is identical to the standard real line at finite temperature.}

By analytic continuation from Euclidean to Lorentzian time, we will also be able to compute the time evolution of entanglement asymmetry in those geometries. Thus entanglement asymmetry can be used to study dynamics: starting with a state that explicitly breaks the symmetry (in our setup regarded as a local quench at $t=0$) we can study the dynamical restoration of the symmetry as expected from thermalization. In the context of quantum spin chains, entanglement asymmetry was used in \cite{Ares:2022koq} and subsequent works to provide a quantitative framework to study the Mpemba effect \cite{Mpemba:1969, Lu:2017, Kumar:2020} --- the observation that states father away from equilibrium can relax faster. We observe a similar effect for the coherent states in CFTs in general dimensions.

First of all, we can use the computation of Section~\ref{sec: rel entropy} of the entanglement asymmetry (\ref{asymmetry Rindler general}) of a CFT$_d$ in the Rindler geometry to determine the asymmetry of a thermal state in hyperbolic space. We already introduced flat coordinates $\{z_1, z_2, y_{i=1, \dots, d-2}\}$ on $\bR^d$, where $z_1$ is Euclidean time, so that the subsystem $A$ is the half-space $A = \{z_1 = 0, z_2 \geq 0\}$. We also introduced a complex coordinate
\be
z = z_2 + i z_1 = r \, e^{i\tau}
\ee
so that the metric takes the form $ds_\text{Rindler}^2 = dr^2 + r^2 d\tau^2 + dy_{1, \dots, d-2}^2$ with $\tau \cong \tau + 2\pi$. Without loss of generality, we took the insertions at $z_\pm$ with $y_i = 0$, as in (\ref{more general z_+-}). We now perform a conformal (Weyl) transformation to $S^1 \times \bH^{d-1}$:
\be
\label{metric hyper r y}
ds_\text{hyper}^2 = d\tau^2 + \frac{dr^2 + dy_{1, \dots, d-2}^2}{r^2} = \frac1{r^2} \, ds_\text{Rindler}^2 \,.
\ee
This maps the subsystem $A$ of the Rindler geometry to the whole hyperbolic space $\bH^{d-1}$ at $\tau=0$. Since entanglement asymmetry transforms as a two-point function, it follows that $\Delta S^{(2)}_\text{hyper} = r^{2\Delta} \Delta S^{(2)}_\text{Rindler}$ and therefore
\be
\label{asymmetry thermal hyperbolic}
\Delta S^{(2)}_\text{hyper} = I_\Delta \Bigl( \tan\bigl( \tfrac{\tau_0}2 \bigr) \Bigr)
\ee
where $\tau_0 = \frac12(\tau_- + \tau_+)$ and $I_\Delta$ is as in (\ref{def function I}). This is the entanglement asymmetry of an excited thermal state (with inverse temperature $\beta = 2\pi$) on the hyperbolic plane. It only depends on $\tau_0$ (related to the Euclidean time of the insertions with respect to the cut) and not on the location on the hyperbolic plane because of its $SO(d-1,1)$ isometry.

It is convenient to rewrite the metric on the hyperbolic geometry as
\be
\label{metric hyper u theta}
ds^2_\text{hyper} = d\tau^2 + du^2 + \sinh^2(u) \, d\Omega_{d-2}^2 = \frac{dz \, d\bar z}{|z|^2} + \frac14 \Bigl\lvert z - \frac1{z^*} \Bigr\rvert^2 d\Omega_{d-2}^2
\ee
where $z = e^{u + i \tau}$ is some other complex coordinate,%
\footnote{The coordinate change from (\ref{metric hyper r y}) to (\ref{metric hyper u theta}) is $(1+r^2 + \vec y^{\,2})/2r = \cosh(u)$, $(1-r^2 - \vec y^{\,2})/2r = \sinh(u) \, \theta_{d-1}$, $y^j/r = \sinh(u) \, \theta_j$ for $j=1, \ldots, d-2$, so that $\theta_{a=1, \ldots, d-1} \in S^{d-2}$ satisfy $\sum_a \theta_a^2 = 1$.}
while $d\Omega_{d-2}^2$ is the angular metric on a unit sphere $S^{d-2}$. Notice that in $d>2$ we take $u\geq 0$ and so $|z| \geq 1$, while in $d=2$ we can neglect $d\Omega_{d-2}^2$ but take $u \in \bR$. We consider a conformal (holomorphic) transformation
\be
z = f(w) \qquad\text{where}\qquad w = r + it_\text{E}
\ee
and $t_\text{E}$ is Euclidean time. The metric takes the form
\be
\label{transformation of metric}
ds^2_\text{hyper} = \biggl\lvert \frac{f'(w)}{f(w)} \biggr\rvert^2 \biggl[ dw\, d\bar w + \frac{ \bigl( |f|^2 -1 \bigr)^2}{4|f'|^2} \, d\Omega_{d-2}^2 \biggr] \,\equiv\, \Omega^2 \, ds^2_\text{geom} \,.
\ee
This allows us to obtain the asymmetry in other geometries. We will be interested in transformations such that the coefficient of $d\Omega_{d-2}^2$ is only a function of $r$ and not of $t_\text{E}$. In any case, the conformal factor $\Omega$ and the parameter $x$ are given by
\be
\label{Omega and x general}
\Omega = \biggl\lvert \frac{f'(w_-)}{ f(w_-)} \biggr\rvert \,,\qquad\qquad
x \equiv \tan\Bigl( \frac{\tau_0}2 \Bigr) = - i \; \frac{ z_- + |z_-| }{ z_- - |z_-|} = - i \; \frac{f(w_-) + |f(w_-)| }{ f(w_-) - |f(w_-)|} \,.
\ee
We thus obtain the general formula
\be
\label{asymmetry general geometry}
\Delta S^{(2)}_\text{geom} = \Omega^{2\Delta} \, I_\Delta(x) \,.
\ee

\subsection{Asymmetry for a finite subregion}
\label{sec: finite subregion}

The choice
\be
z = f(w) = \frac{w + \ell}{\ell - w}
\ee
in (\ref{transformation of metric}) gives the flat metric $ds_\text{geom}^2 = dt_\text{E}^2 + dr^2 + r^2 d\Omega_{d-2}^2$ on $\bR^d$. In $d>2$ dimensions, the preimage of the hyperbolic plane is the spherical region $A = \bigl\{ t_\text{E} = 0 ,\, r^2 \leq \ell^2 \bigr\}$ of radius $\ell$ that we will call a disk. In $d=2$ it is convenient to take $r \in \bR$ and then the subsystem $A$ is the finite interval $r \in [-\ell, \ell]$.
Using (\ref{Omega and x general})--(\ref{asymmetry general geometry}) we can thus compute the entanglement asymmetry of a finite spherical subregion. The conformal factor and the variable $x$ in this case read:
\be
\label{Omega and x inf vol}
\Omega = \frac{2\ell}{ \bigl\lvert w_-^2 - \ell^2 \bigr\rvert} \,,\qquad\qquad x = \frac{2\ell \im(w_-) }{ \ell^2 - |w_-|^2 - | w_-^2 - \ell^2| } \,,
\ee
in terms of the insertion point $w_-$, and $\Delta S^{(2)}_\text{disk} = \Omega^{2\Delta} I_\Delta(x)$ according to (\ref{asymmetry general geometry}).

\begin{figure}[t]
\centering
\includegraphics[width=0.999\textwidth]{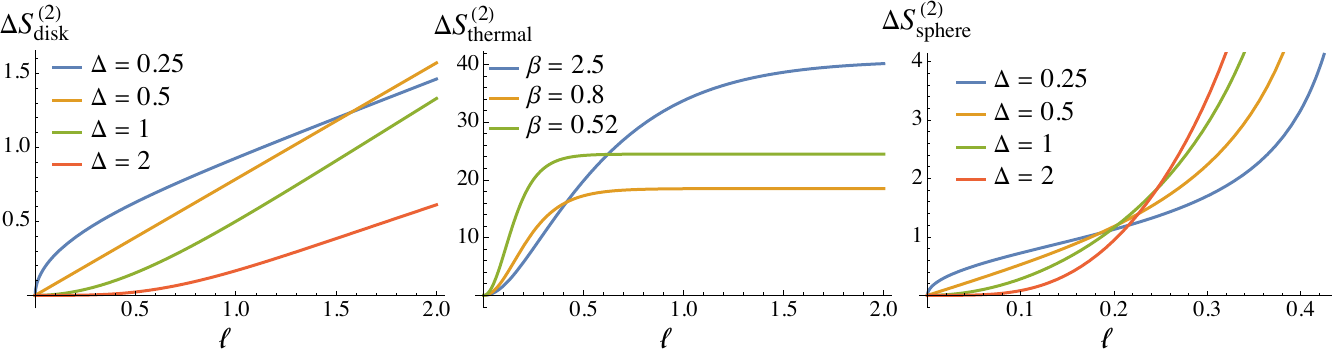}
\caption{\label{fig: finite asymmetries}%
Entanglement asymmetries in various geometries for insertions along the imaginary lines. Left: asymmetry $\Delta S^{(2)}_\text{disk}$ of a finite subregion as function of $\ell$, for $\eta = 1$. Center: asymmetry $\Delta S^{(2)}_\text{thermal}$ at finite temperature $\beta^{-1}$ as function of $\ell$, for $\Delta = 1$ and $\eta = \frac14$. Right: asymmetry $\Delta S^{(2)}_\text{sphere}$ in finite volume as function of $\ell$, for $L =1$ and $\eta = \frac13$ (central line).}
\end{figure}

These formulas simplify if we take the insertion point $w_-$ to lie along the imaginary axis $\re(w)=0$. This is a vertical line (along Euclidean time) that goes through the center of the disk. Let us set%
\footnote{This corresponds to $e^{i\tau_0} = (i\eta - \ell) / (i\eta + \ell)$.}
\be
w_- = - i \eta 
\ee
in terms of a real parameter $\eta \in \bR_+$. Then
\be
\label{Omega and x inf vol im axis}
\Omega = \frac{2 \ell}{\ell^2 + \eta^2} \qquad\qquad\text{and}\qquad\qquad x = \tan\Bigl( \frac{\tau_0}2 \Bigr) = \frac\ell\eta \,.
\ee
We thus obtain the following compact expression for the entanglement asymmetry of a disk in infinite volume:
\be
\label{asymmetry interval}
\Delta S^{(2)}_\text{disk} = \biggl( \frac{2\ell}{\ell^2 + \eta^2} \biggr)^{\! 2\Delta} \, I_\Delta \biggl( \frac\ell\eta \biggr) \,.
\ee
In Appendix~\ref{app: asymmetry from twist} we perform a check of this result by computing it in 2d from a four-point function of $V$, $V^\dag$ and two twist operators (as in \cite{Calabrese:2004eu}), instead of using geometric replicas. In Fig.~\ref{fig: finite asymmetries} (left) we plot the behavior of $\Delta S^{(2)}_\text{disk}$ with $\ell$ for some values of the parameters. The expression in (\ref{asymmetry interval}) has the following asymptotic behaviors:
\be
\ell \ll \eta: \;\; \Delta S^{(2)}_\text{disk} \simeq \frac{\sqrt\pi\; \Gamma(\Delta + 1) }{ 2 \, \Gamma\bigl( \Delta + \frac32 \bigr) } \, \frac{ \ell^{2\Delta} }{ \eta^{4\Delta} } \;,\qquad\qquad
\eta \ll \ell: \;\; \Delta S^{(2)}_\text{disk} \simeq \frac{\pi \Delta}{4^\Delta} \, \frac{\ell}{\eta^{2\Delta + 1}} \;.
\ee
In particular $\Delta S^{(2)}$ increases and diverges as $\ell$ for large intervals. We believe that this behavior will be corrected asymptotically by higher-order contributions to $\Delta S$. Notice also that $\Delta S^{(2)}$ is dimensionful, in accord with the fact that $\Delta S = \lambda^2 \Delta S^{(2)} + O(\lambda^4)$ and $\lambda$ has dimensions of length$^\Delta$. A natural normalization (that we will use later on) would be to fix $\lambda$ in units of $\eta^\Delta$.

\subsection{Asymmetry at finite temperature and in finite volume}
\label{sec: temp and size}

The choice
\be
\label{holo map finite temp}
z = f(w) = \frac{ \sinh\bigl( \pi (w + \ell)/ \beta \bigr) }{ \sinh\bigl( \pi (\ell - w)/\beta \bigr) } \,,
\ee
which implies the identification $w \cong w + i \beta$, when plugged in (\ref{transformation of metric}) gives
\be
ds^2_\text{geom} = dt_\text{E}^2 + dr^2 + \frac{\beta^2}{4\pi^2} \sinh^2\Bigl( \frac{ 2\pi r}\beta \Bigr) \, d\Omega_{d-2}^2 \,.
\ee
This is the metric on $S^1_\beta \times \bH^{d-1}$, where the Euclidean time circle has radius $\beta$. In $d>2$ dimensions we take $r \geq 0$ and the preimage of the hyperbolic plane is the spherical region $A = \{ t_\text{E} = 0, r^2 \leq \ell^2\}$ inside $\bH^{d-1}$. This geometry therefore describes an excited thermal state (with temperature $\beta^{-1}$) on the hyperbolic plane $\bH^{d-1}$ (with curvature proportional to $\beta^{-2}$), reduced to a spherical subsystem $A$. Since the state lives on the hyperbolic plane, it does not originate from the standard thermal vacuum. On the other hand, in $d=2$ dimensions we take $r \in \bR$ and the geometry is just $S^1_\beta \times \bR$ while the subsystem $A$ is the interval $r \in [-\ell, \ell]$: this is a finite subsystem in the standard thermal setup. The entanglement asymmetry is given by (\ref{Omega and x general})--(\ref{asymmetry general geometry}).

For insertions along the imaginary axis $w_- = - i \eta$ (we take $0 < \eta < \beta/2$) namely along the Euclidean circle that goes through the center of the disk $A$, we find
\be
\Delta S^{(2)}_\text{thermal} = \biggl( \frac{2\pi \beta^{-1} \sinh(2\pi \ell/\beta) }{ \cosh(2\pi \ell / \beta) - \cos(2\pi \eta/\beta) } \biggr)^{\! 2\Delta} \, I_\Delta \biggl( \frac{\tanh(\pi\ell/\beta) }{ \tan(\pi\eta/\beta)} \biggr) \,.
\ee
We plot it in Fig.~\ref{fig: finite asymmetries} (center) for some values of the parameters.
In the limit $\beta \gg \ell,\eta$ we recover the case (\ref{asymmetry interval}) of a disk in infinite volume. For $\ell \gg \beta$ we have $\tau_0 \simeq \pi(1-2\eta/\beta)$ and therefore the asymmetry asymptotes to a constant:
\be
\Delta S^{(2)}_\text{thermal} \,\simeq\, (2\pi/\beta)^{2\Delta} I_\Delta\bigl( \cot(\pi \eta/\beta) \bigr) \qquad\qquad\text{for } \ell \gg \beta \,.
\ee
This reproduces (\ref{asymmetry thermal hyperbolic}) for $\beta = 2\pi$.

\bigskip

The choice
\be
\label{holo map finite size}
z = f(w) = \frac{ \sin\bigl( \pi (w+\ell) / L \bigr) }{ \sin\bigl( \pi (\ell - w)/L \bigr) }
\ee
with $0 < \ell < L/2$ in (\ref{transformation of metric}) gives
\be
ds^2_\text{geom} = dt_\text{E}^2 + dr^2 + \frac{L^2}{4\pi^2} \sin\Bigl( \frac{ 2\pi r}L \Bigr)^2 \, d\Omega_{d-2}^2 \;.
\ee
This is the metric on $\bR \times S^{d-1}$ where the sphere has circumference $L$. Notice that (\ref{holo map finite size}) implies the identification $w \cong w + L$. In $d>2$ dimensions we take $0 \leq r \leq \frac L2$ and the preimage of the hyperbolic plane is the spherical region $A = \{ t_\text{E} = 0, r^2 \leq \ell^2\}$ inside the sphere $S^{d-1}$. In $d=2$ we take $r \cong r + L$ and $A$ is the interval $r \in [-\ell, \ell]$ inside this circle. This geometry allows us to evaluate the asymmetry of a spherical disk in finite volume. The general formula is (\ref{Omega and x general})--(\ref{asymmetry general geometry}).

We consider two interesting cases in which the expressions simplify. One is of insertions along the imaginary axis $w_- = - i\eta$ which is the Euclidean time line through the center of the disk. We call this the central line, and we find
\be
\Delta S^{(2)}_\text{sphere} = \biggl( \frac{2\pi L^{-1} \sin(2\pi \ell/L) }{ \cosh(2\pi \eta/L) - \cos(2\pi \ell/L)} \biggr)^{\! 2\Delta} \, I_\Delta \biggl( \frac{\tan(\pi \ell/L) }{ \tanh(\pi \eta/L)} \biggr) \,.
\ee
We plot it in Fig.~\ref{fig: finite asymmetries} (right) for some values of the parameters.
Notice that $2\pi \ell/L < \tau_0 < \pi$. For $L \gg \ell, \eta$ we recover the asymmetry (\ref{asymmetry interval}) of a disk in infinite volume. For $\ell \to L/2$ (\ie, if the subspace $A$ is the whole space) the asymmetry $\Delta S^{(2)}$ diverges, however we believe that this is an artefact of the perturbative expansion. On the other hand, for $\eta \to \infty$ the asymmetry vanishes because the Euclidean time evolution projects the state to the ground state, which is symmetric.

The other case is of insertions along the axis $\re(w) = \frac L2$ that we parametrize as $w_- = \frac L2 -i\eta$. This is the Euclidean time line that goes through the antipodal point on the sphere with respect to the center of the disk, and we call it the antipodal line. We find
\be
\Omega = \frac{2\pi L^{-1} \sin(2\pi \ell/L) }{ \cosh(2\pi \eta/L) + \cos(2\pi \ell/L)} \,,\qquad
x = \tan\Bigl( \frac{\tau_0}2 \Bigr) = \tan(\pi \ell/L) \tanh(\pi \eta/L) \,,
\ee
and $\Delta S^{(2)}_\text{sphere} = \Omega^{2\Delta} I_\Delta(x)$. In this case $0 < \tau_0 < 2\pi\ell/L$.

\subsection{Time dependence and relaxation}

\begin{figure}[t]
\centering
\begin{tikzpicture}
	\draw [black!40!white, ->] (-2.5, 0) -- (2.5, 0);
	\draw [black!40!white, ->] (0, -1.4) -- (0, 1.4);
	\draw [black!70!white] (2.15, 1.4) -- (2.15, 1.1) -- (2.5, 1.1); \node [black!70!white] at (2.35, 1.25) {\small$w$};
	\draw [red!80!black, thick, double] (-1.2, 0.3)  -- (1.2, 0.3) node [pos = 0.1, black, shift={(0,-0.33)}] {\small$A$};
	\filldraw (-1.2, 0.3) circle [radius = 0.05] node [above, shift = {(-0.2, 0)}, black] {\small$-\ell + i\xi$};
	\filldraw (1.2, 0.3) circle [radius = 0.05] node [above, shift = {(0.3, 0)}, black] {\small$\ell + i\xi$};
	\filldraw [blue] (0, -0.8) circle [radius = 0.04] node [right, black] {\small$w_- = -i\eta$};
	\draw [blue, thick] (0, 0.8) circle [radius = 0.04] node [right, black] {\small$w_+$};
	\draw [blue!80!black, dashed] (0, -1.2) -- (0, 1.2);
	\draw [thick, double, ->] (4.5, 0.8) -- (5.3, 0.8) node [pos=0.5, below, shift = {(0, -0.3)}, text width = 2cm, align = center] {\footnotesize analytic continuation \\[-0.1em] $\xi = it$};
\begin{scope}[shift={(9,0)}]
	\draw [black!40!white, ->] (-1.8, 0) -- (1.8, 0);
	\draw [black!40!white, ->] (-135: 1) -- (45: 1);
	\draw [black!70!white] (1.7, 0.7) -- ++(-135: 0.3) -- ++(0.4, 0); \node [black!70!white] at (1.9, 0.63) {\small$w$};
	\draw [red!80!black, thick, double] (-0.9, 1) -- (0.9, 1);
	\draw [dashed] (-0.9, 0) -- (-0.9, 1);
	\draw [dashed] (0.9, 0) -- (0.9, 1);
	\filldraw (-0.9, 1) circle [radius = 0.05];
	\filldraw (0.9, 1) circle [radius = 0.05];
	\draw (-0.9, 0) circle [radius = 0.04];
	\draw (0.9, 0) circle [radius = 0.04];
	\filldraw [blue] (-135: 0.7) circle [radius = 0.04] node [left, black] {\small$w_-$};
	\draw [blue, thick] (45: 0.7) circle [radius = 0.04] node [left, black] {\small$w_+$};
\end{scope}
\end{tikzpicture}
\caption{\label{fig: time shift}%
The time evolution of entanglement asymmetry can be computed by first shifting the cut $A$ along the Euclidean time direction by $\xi$, and then performing analytic continuation $\xi = it$ to Lorentzian time $t$.}
\end{figure}
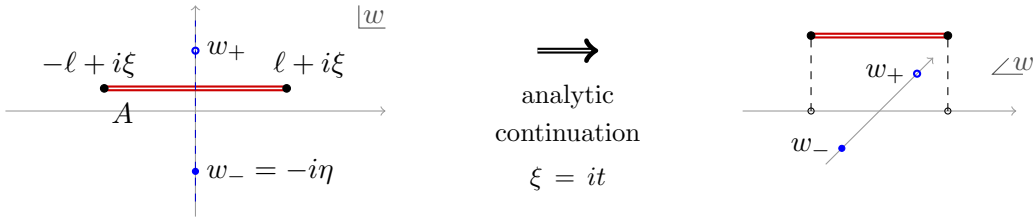

As in \cite{Calabrese:2006rx, Cardy:2016fqc} we can use conformal transformations and analytic continuation to obtain the time evolution of entanglement asymmetry, as follows.
One first computes the entanglement asymmetry of a configuration in which the cut $A$ is shifted by $\xi$ along the Euclidean time direction, as in Fig~\ref{fig: time shift}. In our setup, this is equivalent to a geometry in which $A$ is left untouched at $t_\text{E} = 0$ but the insertions are shifted as $\cO(w_- -i\xi)$ and $\cO^\dag(w_-^* - i \xi)$. Since such a configuration is not specular with respect to the spatial slice of the cut at $t_\text{E} = 0$, the path integral prepares a ``density matrix''
\be
\label{operator in local quench}
\rho_\xi = \Tr_B \Bigl( e^{- (\eta + \xi) H} \, \cO \, |0\rangle \langle 0| \, \cO^\dag \, e^{-(\eta - \xi) H} \Bigr)
\ee
(where $\eta = |\im w_-|$) which, for real $\xi$, is not Hermitian. In order to compute the asymmetry of $\rho_\xi$ in general one needs to extend the evaluation of the integral in (\ref{integral with trace}) (that computes the asymmetry in the Rindler geometry) to the case that $z_\pm$ have different absolute values. In the thermal hyperbolic geometry, this corresponds to having $\cO, \cO^\dag$ inserted at different points on the hyperbolic plane.
Then one takes the analytic continuation $\xi = it$, which produces the time-evolved Hermitian density matrix
\be
\rho(t) = e^{-it H_A} \, \rho \, e^{itH_A} \,.
\ee
Now the configuration is specular with respect to the spatial slice, the cut is shifted in the Lorentzian time direction, and the path integral is performed along a Schwinger--Keldysh contour. This setup can be regarded as a local quench:%
\footnote{We call the quench ``local'' (as opposed to ``global'') because it is localized in space and it produces an inhomogeneous state.}
at $t=0$ the system is in an excited state prepared by the Euclidean path integral, and for $t>0$ is relaxes towards the vacuum.

In the special cases that the orbits of $w_- - i \xi$ and $w_+ - i\xi$ (as we vary $\xi$) have $|z_-| = |z_+|$ in the Rindler geometry (or equivalently $u_- = u_+$ in the thermal hyperbolic geometry), we can use (\ref{asymmetry integral}). These special cases are precisely the ``imaginary lines'' we already discussed in the previous sections. Let us discuss the various cases separately.

\begin{figure}[t]
\centering
\includegraphics[height=4.2cm]{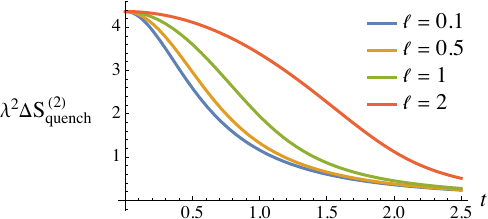}
\caption{\label{fig: quench}%
Time evolution of the entanglement asymmetry $\lambda^2 \Delta S^{(2)}_\text{quench}(t)$ of a disk after a local quench, for $\eta = 0.6$, $\Delta = 0.5$, and different values of the disk radius $\ell$. We normalized $\lambda^2 \propto \ell^{-1}$ in order to have the same initial value. Notice that larger subsystems thermalize more slowly.}
\end{figure}

\paragraph{Finite spherical subregion.}
We consider the geometry of Section~\ref{sec: finite subregion} with a spherical subsystem $A = \{t_\text{E} = 0, r^2 \leq \ell^2\}$ in $\bR^d$ and insertions along the imaginary axis $\re(w) =0$. With insertions at $w_- = - i (\eta + \xi)$ and $w_+ = i(\eta - \xi)$ we find
\be
\tan\Bigl( \frac{\tau_-}2 \Bigr) = \frac{\ell}{\eta + \xi} \;,\qquad\qquad \tan\Bigl( \frac{\tau_+}2 \Bigr) = \frac{\ell}{\eta - \xi} \;.
\ee
Including the conformal factors, the entanglement asymmetry takes the form:
\be
\Delta S^{(2)} = \biggl[ \frac{4\ell^2}{ \bigl( \ell^2 + (\eta+\tau)^2 \bigr) \bigl( \ell^2 + (\eta - \tau)^2 \bigr) } \biggr]^\Delta \, I_\Delta\Bigl( \tan^2\bigl( \tfrac{\tau_0}2 \bigr) \Bigr)
\ee
where $\tau_0 = \frac12 (\tau_- + \tau_+)$. We then perform the analytic continuation $\xi = it$ and obtain
\be
\label{Delta S quench}
\Delta S^{(2)}_\text{quench} = \biggl\lvert \frac{ 2\ell }{ \ell^2 + (\eta + it)^2 } \biggr\rvert^{2\Delta} \, I_\Delta(x)
\ee
where
\be
x = \tan \biggl[ \re \arctan\biggl( \frac{ \ell }{ \eta + it} \biggr) \biggr] = \frac{ \ell^2 - \eta^2 - t^2 + \sqrt{ \bigl( \ell^2 - \eta^2 - t^2 \bigr)^2 + 4\ell^2 \eta^2 } }{ 2 \ell \eta} \,.
\ee
Both the conformal factor $\Omega$ and $x$ are as in (\ref{Omega and x inf vol}) but evaluated at $w_- = t -i \eta$. This describes the time evolution of entanglement asymmetry after a local quench. We plot $\lambda^2 \Delta S^{(2)}_\text{quench}(t)$ for a choice of $\Delta, \eta$ and different values of $\ell$ in Fig.~\ref{fig: quench}. Using a normalization of $\lambda^2$ so as to have the same initial value for the asymmetry, we see that larger subsystems thermalize more slowly. We will explore other physical consequences of (\ref{Delta S quench}) in Section~\ref{sec: Mpemba effect}.

\paragraph{Finite temperature.} For the finite-temperature geometry $S^1_\beta \times \bH^{d-1}$ of Section~\ref{sec: temp and size} and insertions along the imaginary axis, proceeding as before we find
\be
\label{Delta S T-quench}
\Delta S^{(2)}_\text{T quench} = \biggl\lvert \frac{ 2\pi \beta^{-1} \sinh(2\pi \ell/\beta) }{ \cosh(2\pi \ell/\beta) - \cos \bigl( 2\pi (\eta + it) / \beta \bigr) } \biggr\rvert^{2\Delta} I_\Delta(x)
\ee
with
\be
x = \tan \biggl[ \re \arctan\biggl( \frac{\tanh( \pi \ell /\beta) }{ \tan\bigl( \pi( \eta + it) / \beta \bigr) } \biggr) \biggr] \,.
\ee
The parameter $x$ can also be expressed as in (\ref{Omega and x general}) with $w_- = t -i \eta$.

\paragraph{Finite volume.} For the geometry $\bR \times S^{d-1}$ of Section~\ref{sec: temp and size} and insertions along the imaginary axis (that we called the central line), we find
\be
\label{Delta S L-quench}
\Delta S^{(2)}_\text{L quench} = \biggl\lvert \frac{ 2\pi L^{-1} \sin(2\pi \ell/L) }{ \cosh \bigl(2\pi (\eta + it)/L \bigr) - \cos( 2\pi \ell / L) } \biggr\rvert^{2\Delta} I_\Delta(x)
\ee
with
\be
x = \tan \biggl[ \re \arctan\biggl( \frac{\tan( \pi \ell / L) }{ \tanh\bigl( \pi( \eta + it) / L \bigr) } \biggr) \biggr] \,.
\ee
One can obtain a similar expression for insertions along the antipodal line. In both cases the parameter $x$ can also be expressed as in (\ref{Omega and x general}) with $w_- = t -i \eta$.

\subsection{Mpemba effect}
\label{sec: Mpemba effect}

In this section we analyze some physical consequences of the time evolution of entanglement asymmetry. Before doing that, notice that the primary operator $V$ has dimensions of mass$^\Delta$ while $\lambda$ has dimensions of length$^\Delta$. It is then more natural to parametrize
\be
\lambda = \epsilon \, \eta^\Delta
\ee
in terms of a (small) dimensionless parameter $\epsilon$. This is the normalization we will use in the rest of this section, hence the entanglement asymmetry takes the form
\be
\Delta S = \epsilon^2 \, \eta^{2\Delta} \, \Omega^{2\Delta} \, I_\Delta(x) + O(\epsilon^4)
\ee
at leading order in $\epsilon$.

\begin{figure}[t]
\centering
\includegraphics[height = 4.2cm]{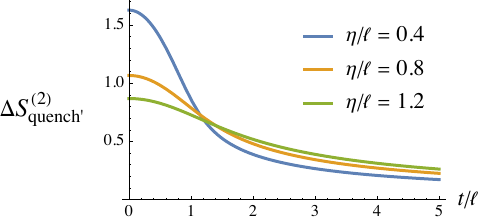}
\caption{\label{Fig:plotM}%
Thermalization curves $\Delta S^{(2)}_{\text{quench}'}(t) \,\equiv\, \eta^{2\Delta} \Delta S^{(2)}_\text{quench}(t)$ for states created with different values of $\eta$ at fixed $\Delta = 0.2$. We observe the Mpemba effect between any two such curves.}
\end{figure}

\begin{figure}[t]
\centering
\includegraphics[height = 4.2cm]{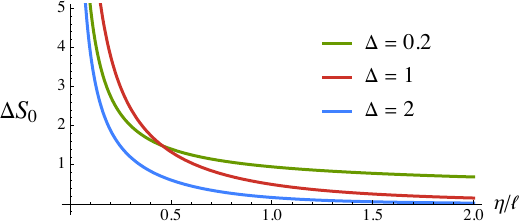}
\caption{\label{Fig:plotInitEta}%
The initial value $\Delta S_0(\eta) = \eta^{2\Delta} \Delta S^{(2)}_\text{disk}$ is a decreasing function of $\eta$ for any fixed $\Delta$. Since the thermalization timescale $t_*$ is an increasing function of $\eta$, we always observe the Mpemba effect among states of different $\eta$ and same $\Delta$, as illustrated in Fig.~\ref{Fig:plotM}.}
\end{figure}

Let us first study the case of a finite subsystem in infinite volume given by (\ref{Delta S quench}). Since the conformal factor and the parameter $x$ behave as $\Omega \sim 2\ell/t^2$ and $x \sim \eta\ell/t^2$ for large $t$, at late times the asymmetry is controlled by the constant $I_\Delta(0)$ and more precisely
\be
\Delta S \,\sim\, \epsilon^2 \, \frac{\sqrt\pi \, \Gamma(\Delta+1) }{ 2 \, \Gamma(\Delta + 3/2)} \, \biggl( \frac{\eta\ell}{t^2} \biggr)^{\! 2\Delta} \qquad\qquad \text{for } t \to \infty \,.
\ee
Defining a thermalization timescale $t_\ast$ as the time at which $\Delta S$ gets reduced by a factor of $e$, we obtain
\be
t_* = \sqrt{\eta \ell} \; e^{1/4\Delta} \,.
\ee
We see that thermalization is slower if we increase $\eta$ while it is faster if we increase the conformal dimension $\Delta$. The two parameters $\eta,\Delta$ label the class of states we consider, and it is interesting to compare the thermalization of different states.

Consider two symmetry-breaking states $\rho$, $\rho'$ and compare how fast they thermalize. How much these states break the symmetry is quantified by the initial value $\Delta S_0$ of the asymmetry at $t=0$.  The speed at which they thermalize is quantified by the corresponding thermalization timescales $t_*^{\phantom\prime}$ and $t_\ast'$. The Mpemba effect \cite{Mpemba:1969, Lu:2017, Kumar:2020} occurs when, despite the symmetry being more broken, the thermalization timescale is shorter. This can be expressed using a quantity
\be
m = \frac{\Delta S_0' - \Delta S_0^{\phantom\prime} }{ t'_* - t_*^{\phantom\prime} } \,.
\ee
If $m\geq 0$ the state starting at a higher $\Delta S_0$ takes more time to equilibrate and there is no Mpemba effect. If $m<0$, instead, that state relaxes faster, the curves cross, and the Mpemba effect takes place.

If we consider a one-parameter family of states, $m$ evaluated for infinitesimally closed states is related to the derivative of $\Delta S_0$ with respect to the parameter. The sign of the derivative governs whether the Mpemba effect occurs locally around a given state. The result is that the Mpemba effect occurs only in some regions of parameter space. To illustrate this point we analyze two cases: varying $\eta$ at fixed $\Delta$, and varying $\Delta$ at fixed $\eta$. In the first case we always observe the Mpemba effect, while in the second case we only observe it using subsystems with $\ell > \eta$ and for sufficiently small conformal dimensions.

\begin{figure}[t]
\centering
\subf{\includegraphics[height = 4.7cm]{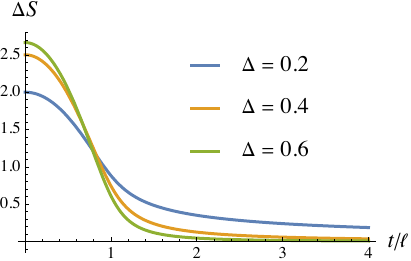}}{$\eta/\ell = 0.3$}
\hspace{2em}
\subf{\includegraphics[height = 4.7cm]{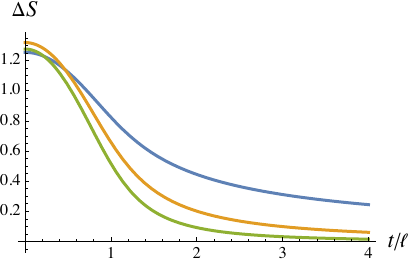}}{$\eta/\ell = 0.6$} \\
\centering
\subf{\includegraphics[height = 4.7cm]{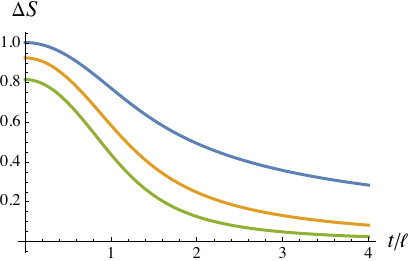}}{$\eta/\ell = 0.9$}
\hspace{2em}
\subf{\includegraphics[height = 4.7cm]{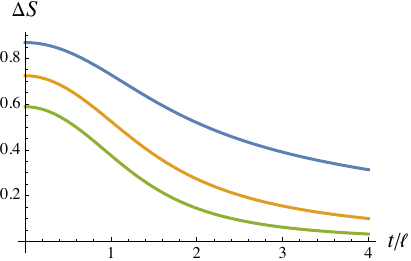}}{$\eta/\ell =1.2$}
\caption{\label{Fig:plotMDelta}%
Thermalization for different values of $\Delta$ and $\eta$. We observe different behaviors in different regions of parameter space. The Mpemba effect is only visible in the two upper plots.}
\end{figure}

\begin{figure}[t]
\centering
\includegraphics[height = 4.7cm]{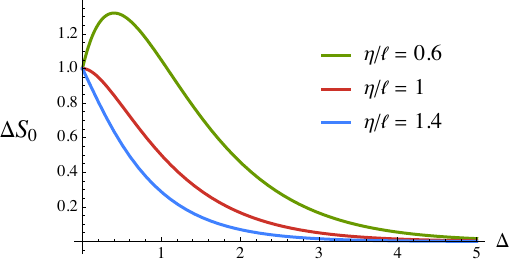}
\caption{\label{Fig:plotInitDelta}%
The curve $\Delta S_0(\Delta) = \eta^{2\Delta} \Delta S^{(2)}_\text{disk}$ has two different shapes depending on the value of the parameter $\eta$. For $\eta \geq \ell$ it is decreasing, while for $\eta < \ell$ it first increases and then decreases.}
\end{figure}

\paragraph{Varying \matht{\eta} at fixed \matht{\Delta}.} This is illustrated in Fig.~\ref{Fig:plotM}. We always observe the Mpemba effect because any two curves cross.  This can be understood from the fact that the initial value $\Delta S_0(\eta)$ is a decreasing function of $\eta$ (see Fig.~\ref{Fig:plotInitEta}) and that the thermalization timescale $t_*$ is an increasing function of $\eta$.

\paragraph{Varying \matht{\Delta} at fixed \matht{\eta}.} As illustrated in Fig.~\ref{Fig:plotMDelta}, we observe two different behaviors depending on the values of the parameters. The two regimes originate from two different profiles of the initial-value curve $\Delta S_0(\Delta)$ plotted in Fig.~\ref{Fig:plotInitDelta}.
If the dimension $\Delta$ is above a critical value $\Delta_*$, then the Mpemba effect does not take place. If $\Delta < \Delta_*$ then there is Mpemba effect, but this is only visible in subsystems $A$ with $\ell > \eta$. This is because at least one of the two conformal dimensions must be below the value $\Delta_*$ at which $\Delta S_0$ has a maximum: The condition to observe the Mpemba effect is that $\Delta < \Delta'$ with $\Delta S_0(\Delta) <  \Delta S_0(\Delta')$ which is possible only if $\Delta < \Delta_*$ and only in the region $\ell > \eta$.

\begin{figure}[t]
\centering
\includegraphics[width=0.999\textwidth]{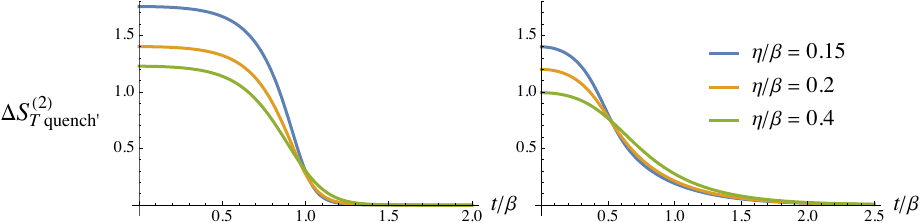}
\caption{\label{Fig:plotTemp}%
Relaxation of entanglement asymmetry $\Delta S^{(2)}_{\text{T quench}'}(t) = \eta^{2\Delta} \Delta S^{(2)}_\text{T quench}(t)$ at finite temperature $\beta^{-1}$, for different values of the parameters. Left plot: $\Delta = 1$ and $\ell/\beta =1$. Right plot: $\Delta = 0.2$ and $\ell/\beta = \frac12$.}
\end{figure}

\subsubsection*{Finite temperature or volume}

\begin{figure}[t]
\centering
\includegraphics[height = 4.3cm]{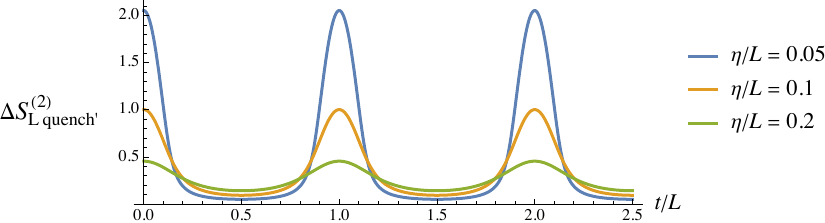}
\caption{\label{Fig:plotSize}%
Thermalization in finite volume described by $\Delta S^{(2)}_{\text{L quench}'}(t) = \eta^{2\Delta} \Delta S^{(2)}_\text{L quench}(t)$ for different values of $\eta$. We chose $\Delta =1$ and subsystem size $\ell = L/8$. We observe a periodic behavior with period $L$.}
\end{figure}

Next, we can study the finite-temperature case given by $\Delta S^{(2)}_\text{T quench}(t)$ in (\ref{Delta S T-quench}). At late times the conformal factor behaves as $\Omega \sim 4 \pi T \sinh(2\pi T\ell) \, e^{-2\pi T t}$ where $T = \beta^{-1}$ is the temperature, while $x \to 0$. This means that at non-vanishing temperature the late-time asymmetry decays exponentially:
\be
\Delta S \,\sim\, \epsilon^2 \, \frac{\sqrt\pi \, \Gamma(\Delta+1) }{ 2\, \Gamma(\Delta + 3/2)} \, \bigl( 2\pi T \eta \sinh(2\pi T \ell) \bigr)^{2\Delta} \, e^{- 4\pi \Delta T t} \qquad\text{for } t \to \infty \,.
\ee
The natural thermalization timescale is
\be
t_* = \frac1{\Delta T} \,.
\ee
The entanglement asymmetry is approximately constant for small times, and then it decays exponentially as illustrated in Fig.~\ref{Fig:plotTemp}. For what regards the Mpemba effect, there are four possible regimes depending on the shape of the curve $\Delta S_0(\Delta)$, distinguished by whether $\lim_{\Delta \to \infty} \Delta S_0$ vanishes or diverges, and by the sign of $\partial_\Delta \Delta S_0 |_{\Delta=0}$. The four phases are:
\begin{itemize}[noitemsep, topsep=0pt]
\item $\Delta S_0$ is a monotonously decreasing function of $\Delta$: we never see the Mpemba effect.
\item $\Delta S_0$ first increases and then it decays with $\Delta$: we observe the Mpemba effect for sufficiently small conformal dimensions.
\item $\Delta S_0$ is monotonously increasing with $\Delta$: we always observe the Mpemba effect.
\item $\Delta S_0$ is first decreasing and then diverging with $\Delta$: we see the Mpemba effect for sufficiently large conformal dimensions.
\end{itemize}

Finally, let us briefly study the time evolution of entanglement asymmetry in finite volume. In Fig.~\ref{Fig:plotSize} we plot the time dependence of $\Delta S^{(2)}_\text{L quench}(t)$ in (\ref{Delta S L-quench}) for different values of the parameter $\eta$ (that controls insertions along the central line) and a choice of $\Delta$. We observe that the asymmetry is periodic with period $L$ (the circumference of the sphere) and thus it never thermalizes. Although oscillations and partial revivals are expected in systems in finite volume (see for instance \cite{Cardy:2014rqa} for a discussion in two-dimensional rational CFTs), we do not expect exact periodicity in a generic CFT. We believe that this is an artefact of the perturbative expansion, that will be corrected by higher-order terms.


\section{Asymmetry in holography}
\label{sec: holography}

The AdS/CFT correspondence  \cite{Maldacena:1997re, Gubser:1998bc, Witten:1998qj} provides a non-perturbative definition of quantum gravity in AdS space in terms of a CFT that lives at the boundary, such that the bulk direction emerges from the strongly-coupled many-body dynamics of the CFT. Quantum information theory has long been recognized to play a primary role in understanding the emergence of the bulk. The Ryu--Takayanagi formula \cite{Ryu:2006bv} shows that complicated quantum-information quantities in the CFT can translate to simple geometric observables in AdS, which sometimes can be derived using a gravitational version of the replica trick \cite{Lewkowycz:2013nqa, Penington:2019kki, Almheiri:2019qdq}. Thus, it is natural to expect that entanglement asymmetry is also dual to a simple quantity in the bulk. In this section we provide some evidence for this, at least at leading order in the parameter $\lambda$.

\subsection{Holographic setup}
\label{sec: holographic setup}

A holographic CFT$_d$ on $\bR^d$ is dual to a bulk gravitational theory in Euclidean AdS$_{d+1}$ with metric (in Poincar\'e coordinates):
\be
ds^2 = \frac{dz^2+ dx_1^2 + \ldots + dx_d^2 }{ z^2} \,.
\ee
On the boundary we consider states obtained by turning on a source for a charged primary operator $V$ in the Euclidean past:
\be
|\psi \rangle = e^{i \!\int\! d^d x \: \lambda(x) \, V(x)} \, |0 \rangle \,.
\ee
Such states are of particular interest in holography because they correspond to classical bulk states (see \cite{Faulkner:2017tkh} and references therein). When $V$ is charged under a $U(1)$ symmetry, the state $|\psi\rangle$ has nontrivial entanglement asymmetry.

The charged primary $V$ of conformal dimension $\Delta$ is dual to a complex scalar field $\phi$ in the bulk, and its source $\lambda(x)$ corresponds to a boundary condition
\be
\label{boundary condition}
\lim_{z\to 0} \, z^{\Delta-d} \, \phi(z, x) = \lambda(x) \,.
\ee
The matter theory in the bulk has Euclidean action:
\be
\label{bulk action}
S = \int\! dz \, d^dx \, \sqrt{g} \, \biggl( \frac14 F_{\mu\nu} F^{\mu\nu} + \bigl\lvert D_\mu \phi \bigr\rvert^2 + m^2 |\phi|^2 \biggr) \,,
\ee
where $D_\mu\phi = \nabla_\mu \phi - i e A_\mu\phi$ and the mass is related to the dimension of $V$ via the formula $m^2 = \Delta (\Delta -d)$.
The equations of motion are
\be
\bigl( D_\mu D^\mu - m^2 \bigr) \, \phi = 0 \,, \qquad \nabla^\nu F_{\mu\nu} = j_\mu \qquad\text{with}\qquad j_\mu = - i \, \bigl( \phi^\dag D_\mu \phi - \phi \, D_\mu \phi^\dag \bigr) \,,
\ee
where $j_\mu$ is the bulk current. The solution that satisfies the boundary condition (\ref{boundary condition}) can be written as \cite{Witten:1998qj, Freedman:1998tz}:
\be\label{eq:phi=K}
\phi(z,x) = \int d^dx' \; K_E(z,x \,|\, x') \, \lambda(x')
\ee
where the bulk-to-boundary propagator is (for $\Delta > d/2$):
\be
\label{bulk-to-boundary prop Poincare}
K_E(z, x \,|\, x') = C_\Delta \, \biggl( \frac{z}{ z^2 + (x - x')^2} \biggr)^\Delta \,.
\ee
The value of $C_\Delta$ that reproduces (\ref{boundary condition}) is $C_\Delta = \Gamma(\Delta) / \bigl( \pi^{d/2} \, \Gamma(\Delta - d/2) \bigr)$.%
\footnote{For the case $\Delta = \frac d2$ in which the mass is at the Breitenlohner--Freedman bound, the asymptotic behavior and the normalizations are different \cite{Freedman:1998tz}.}
However, with that value, the resulting two-point function of the dual complex scalar operator $V$ has coefficient \mbox{$c = 2 \, (2\Delta -d) \, \Gamma(\Delta) / \bigl( \pi^{d/2} \, \Gamma(\Delta - d/2) \bigr)$} (see the Appendix of \cite{Freedman:1998tz}). In the field theory analysis we normalized the operator $V$ so that its two-point function has coefficient 1. In order to have the same normalization in holography, we should rescale the field $\phi$ by $c^{-1/2}$. This is the same as rescaling $K_E$, namely, as taking
\be
C_\Delta = \sqrt{ \frac{ \Gamma(\Delta) }{ 4 \, \pi^{d/2} \, \Gamma\bigl( \Delta + 1 - \frac d2 \bigr) } }
\ee
in (\ref{bulk-to-boundary prop Poincare}). Notice that the unitarity bound guarantees that $\Delta + 1 - \frac d2 > 0$.

We focus on states where the operator is inserted at a single point:
\be
\lambda(x) = \lambda \; \delta^d(x-x_0) \,,
\ee
and we take $\lambda \in \bR$.
We consider spherical regions $A$ in the CFT, such that the modular Hamiltonian is local and related to the boost generators of half-space by a conformal transformation \cite{Casini:2011kv}.  As explained in Section~\ref{sec:conformalmaps}, the reduced density matrix of $A$ can be obtained by using a different time slicing of $\bR^d$. This corresponds to writing the metric of $\bR^d$ as a conformal transformation of $S^1 \times \bH^{d-1}$, namely as
\be
ds^2 =\Omega^{-2} \, \bigl( d\tau^2 + du^2 + \sinh^2(u) \, d\Omega_{d-2}^2 \bigr)
\ee
as we already did in Section~\ref{sec: finite subregion}. The conformal factor (\ref{Omega and x inf vol}) in these coordinates (for $\ell=1$) is
\be
\label{conformal factor from holography}
\Omega = \cosh(u) + \cos(\tau) \,.
\ee
The spherical region $A$ is mapped to the whole hyperbolic plane $\bH^{d-1}$ at $\tau=0$.
In the bulk this slicing means writing the AdS metric using coordinates adapted to the Rindler wedge:
\be
\label{eq:rindlermetric}
ds^2 = (\rho^2-1) \, d\tau^2 + \frac{ d\rho^2 }{ \rho^2- 1 } + \rho^2 \, \bigl( du^2 + \sinh^2(u) \, d\Omega_{d-2}^2 \bigr)
\ee
with $\rho > 1$. In these coordinates the bulk-to-boundary propagator (\ref{bulk-to-boundary prop Poincare}) in the embedding formalism reads
\be
\label{eq:bbhyperbolic}
K_E \bigl( \rho, \tau, Y \bigm| \tau', Y' \bigr) = \frac{C_\Delta }{ \bigl( -2 \rho \, Y \cdot Y' - 2 \sqrt{\rho^2-1} \, \cos(\tau - \tau') \bigr)^\Delta } \,.
\ee
Here $Y = \bigl( Y_0, Y_{a=1, \dots, d-1} \bigr)$ are coordinates on the hyperboloid $\bH^{d-1}$, they solve the constraint $Y \cdot Y \equiv - Y_0^2 + \sum_a Y_a^2 = -1$ and are parametrized as $Y = \bigl( \cosh(u), \sinh(u)\, \theta_a \bigr)$ with $\sum_a \theta_a^2 = 1$ so that $\theta_a \in S^{d-2}$ (see Appendix~\ref{app: embedding} and App.~A of \cite{Faulkner:2017tkh} for more details). The Rindler horizon is at $\rho = 1$ and the asymptotic boundary is at $\rho = \infty$. The prescription of \cite{Faulkner:2017tkh}, as we review below, requires us to construct a real-time solution for the bulk fields. In particular, the bulk solution is constructed by analytically continuing the propagator \eqref{eq:bbhyperbolic} to $\tau \to it$. It is then useful at this stage to perform the analytic continuation $\tau = i t$ to Lorentzian signature in \eqref{eq:rindlermetric} as well, where $t$ is the Lorentzian Rindler time. In these coordinates the bulk modular flow in the Rindler wedge is represented by the manifest Killing symmetry 
\be
\label{eq:modularflow}
\xi = 2\pi \, \partial_t \,.
\ee
The prescription of \cite{Faulkner:2017tkh} then, given a spherical region in the CFT$_d$ on $S^1 \times \bH^{d-1}$ as before and a one-parameter family of excited states $\rho_A = \sigma_A \bigl( 1 + i \!\int\! d^dx \, \lambda(x) \, V(x) + \text{h.c.} + \ldots \bigr)$, requires us to construct bulk solutions consistent with the operator insertions on the boundary. In hyperbolic coordinates, \eqref{eq:phi=K} reads
\be
\phi(\rho, t, Y) = \int\! d\tau' \, dY' \; K_E \bigl( \rho, it, Y \bigm| \tau',Y' \bigr) \, \lambda(\tau',Y')
\ee
with $K_E$ as defined in \eqref{eq:bbhyperbolic} but analytically continued to $\tau \to it$.

Let us construct these solutions explicitly in Lorentzian AdS$_{d+1}$. The boundary CFT has delta-function sources corresponding to the Euclidean insertion of two conjugate operators $V$ at $\tau' = - (\pi - \tau_0)$ and $V^\dag$ at $\tau' = (\pi-\tau_0)$. Without loss of generality we take $u' = 0$. The solution takes the form $\phi(\rho, t, u) = \lambda \, K_E \bigl( \rho, it, u \bigm| \tau_0 - \pi , 0 \bigr)$ and similarly for $\phi^\dag$:
\bea
\label{eq:sol}
\phi = \frac{\lambda \, C_\Delta}{ \bigl( 2 \rho \cosh(u) + 2 \sqrt{\rho^2-1} \, \cos(it - \tau_0) \bigr)^\Delta} \,, \\
\phi^\dag = \frac{ \lambda \, C_\Delta }{ \bigl( 2 \rho \cosh(u) + 2 \sqrt{\rho^2-1} \, \cos(it + \tau_0) \bigr)^\Delta } \,.
\eea
Notice that inserting the boundary operators at specular points $\tau' = \mp (\pi - \tau_0)$ automatically ensures that the bulk solutions are complex conjugate to each other.

\subsection{Asymmetry as canonical energy}

We saw that, at leading order, entanglement asymmetry is equal to perturbative relative entropy. The latter was shown in \cite{Lashkari:2015hha, Faulkner:2017tkh} to be holographically dual to canonical energy, first defined by Hollands and Wald \cite{Hollands:2012sf} and then generalized in \cite{Lashkari:2016idm, May:2018tir}. This defines a metric on the space of perturbations, known as the ``quantum Fisher'' or Bogoliubov--Kubo--Mori metric. We therefore conclude that the leading-order entanglement asymmetry is given by the canonical energy, where we are interested in the particular case of scalar primary operator insertions on the boundary, and corresponding dual bulk scalar fields. The canonical energy is defined as a particular symplectic flux of on-shell fields, evaluated on a codimension-one surface in the entanglement wedge which is bounded by the Ryu--Takayanagi surface $\wt A$ on one side and by the boundary subsystem $A$ on the other side. Since the symplectic flux is conserved, we are free to choose any such surface and we choose the $t=0$ spatial slice of the entanglement wedge.%
\footnote{The choice made in \cite{Faulkner:2017tkh} is instead that $\Sigma$ is the horizon of the entanglement wedge.}
We denote such a surface as $\Sigma$, whose boundary is $\partial\Sigma = A \cup \wt A$. The bulk formula for the entanglement asymmetry is
\be
\label{canonical energy}
\Delta S = \int_\Sigma \star \, {\boldsymbol\omega}(\phi, \cL_\xi \phi) + O(\lambda^4) \,,
\ee
where $\phi$ is taken of order $\lambda$.
Here $\xi$ is the vector field of bulk modular flow, which in Rindler coordinates is \eqref{eq:modularflow}, while $\cL_\xi$ is a Lie derivative. We stress that $A$ should be a half-space, a spherical region, or a conformal transformation thereof, in order for the modular flow $\xi$ to be geometric and the formula (\ref{canonical energy}) to be applicable. The symplectic current ${\boldsymbol\omega}$ is a one-form, quadratic in the on-shell bulk fields \eqref{eq:sol}. We also expect corrections suppressed by $G_N$ and $e$ obtained by taking into account the gravitational and gauge fields.

The symplectic current can be derived by writing the variation of the bulk Lagrangian under a field variation \cite{Lee:1990nz}:
\be
\delta \mathscr{L} = E_\phi \, \delta\phi + \nabla^\mu \Theta_\mu[\delta\phi] \,,
\ee
where $E_\phi$ are the equations of motion. The quantity $\Theta_\mu$ that produces a total derivative is called the pre-symplectic current.
The symplectic current is then defined as \cite{Lee:1990nz}:
\be
{\boldsymbol\omega}_\mu \bigl( \delta_1 \phi, \delta_2 \phi \bigr) = \delta_1 \Theta_\mu [\delta_2\phi] - \delta_2 \Theta_\mu [\delta_1\phi] \,.
\ee
For the bulk action (\ref{bulk action}), and working in the approximation that only the complex scalar field is turned on at leading order, we find:
\bea
\label{sympl}
\Theta_\mu[\delta\phi] &= (D_\mu \phi) \, \delta\phi^\dag + (D_\mu \phi^\dag) \, \delta\phi \,, \\
{\boldsymbol\omega}_\mu \bigl( \phi_1, \phi_2 \bigr) &= (D_\mu \phi_1) \, \phi_2^\dag + (D_\mu \phi_1^\dag ) \, \phi_2 - (D_\mu \phi_2) \, \phi_1^\dag - (D_\mu \phi_2^\dag) \, \phi_1 \,.
\eea
The surface $\Sigma$ is at $t=0$. This surface has normal vector $n = (\rho^2-1)^{-1/2} \, \partial_t$ and induced spatial metric
\be
ds^2 \Big|_{t=0} = \frac{d\rho^2}{\rho^2-1} + \rho^2 \, \Bigl( du^2 + \sinh^2(u) \, d\Omega_{d-2}^2 \Bigr) \,\equiv\, h_{ab} \, dx^a dx^b \,.
\ee
The holographic result for the entanglement asymmetry is then
\be
\Delta S = \int_1^\infty \!\! d\rho \int_0^\infty \!\! du \int_{S^{d-2}} \!\! d\Omega_{d-2} \, \sqrt{h} \, n^\mu {\boldsymbol\omega}_\mu (\phi, \cL_\xi \phi) + O(\lambda^4) \,.
\ee
Inserting the solutions \eqref{eq:sol}, the integrand takes the explicit form%
\footnote{Recall that $\Vol_{S^{d-2}} = 2 \pi^{(d-1)/2} / \, \Gamma\bigl( \frac{d-1}2 \bigr)$.}
\begin{align}
& \int\! d\Omega_{d-2} \, \sqrt{h} \, n^\mu {\boldsymbol\omega}_\mu(\phi, \cL_\xi\phi) \Big|_{t=0} = {} \\
&\qquad = \lambda^2 \, \frac{ C_\Delta^2 \, \Vol_{S^{d-2}} \, \pi \, \Delta \, \rho^{d-1} \, \sinh^{d-2}(u) \, \Bigl( 1 + 2 \Delta \sin^2(\tau_0) + \rho \, (\rho^2-1)^{-\frac12} \cos(\tau_0) \cosh(u) \Bigr) }{ 4^{\Delta-1} \, \bigl( \rho \cosh(u) + \sqrt{\rho^2-1} \, \cos(\tau_0) \bigr)^{2(\Delta+1)} } \,. \quad \nn
\end{align}
Note that the integral in $\rho$ converges if $\Delta$ is larger than the unitarity bound. Although the integral is hard to perform analytically, it is easy to evaluate numerically and we verified that it precisely reproduces the CFT answer:
\be
\int_1^\infty \!\! d\rho \int_0^\infty \!\! du \int \! d\Omega_{d-2} \, \sqrt{h} \, n^\mu {\boldsymbol\omega}_\mu(\phi, \cL_\xi \phi) = \lambda^2 \, \Delta S_\text{hyper}^{(2)}
\ee
as given explicitly by eqn.~\eqref{asymmetry thermal hyperbolic}.

\subsection{Relation to the bulk charge}

In holography a bulk gauge field $A_\mu$ is dual to a boundary current operator. However, there also exists a \emph{bulk} current which, for the action in (\ref{bulk action}), takes the form
\be
j_\mu = - i \, \bigl( \phi^\dag D_\mu\phi  - \phi \, D_\mu\phi^\dag \bigr) \,.
\ee
This is written in Lorentzian signature where Hermitian conjugation acts in the standard way. Notice that the bulk current is a rather mysterious quantity from the point of view of the boundary CFT. A natural observable is the bulk charge of the Rindler wedge:
\be
Q_\text{bulk} = \int_\Sigma  \star \, j \,.
\ee
In AdS$_{d+1}$ it takes the explicit form $Q_\text{bulk}  = \int_1^\infty \! d\rho \int_0^\infty \! du \int\! d\Omega_{d-2} \, \sqrt{h} \, n^\mu j_\mu$ and the integrand reads
\be
\int \! d\Omega_{d-2} \, \sqrt{h} \, n^\mu j_\mu \Big|_{t=0} = - \lambda^2 \, \frac{ C_\Delta^2 \, \Vol_{S^{d-2}} \, 2\Delta \, \rho^{d-1} \sinh^{d-2}(u) \sin(\tau_0) }{ 4^\Delta \, \sqrt{\rho^2-1} \, \bigl( \rho \cosh(u) + \sqrt{\rho^2-1} \, \cos(\tau_0) \bigr)^{2\Delta+1} } \,.
\ee
Using the equation of motion $d \star F = \star \, j$ we can write the integral as
\be
Q_\text{bulk} = \int_A \star \, F - \int_{\wt{A}} \star \, F \,,
\ee
where we used that the boundary of the surface $\Sigma$ is the union of the boundary subsystem $A$ and the Ryu--Takayanagi surface $\wt A$. The first term simply computes the expectation value of the $U(1)$ charge $Q$ in the CFT:
\be
\int_A \star \, F = \Tr( \rho \, Q) \,,
\ee
while the second term is the electric flux through the Ryu--Takayanagi surface. 

Now, one might expect a relation between bulk charge and entanglement asymmetry. Indeed we observe that, in our setup, the following simple relation holds:
\be
\label{final formula from holography}
\lambda^2 \, \Delta S^{(2)} = -2\pi \, \partial_{\tau_0} \, Q_\text{bulk} \,.
\ee
This follows from the fact that the integrand satisfies (for any $t$):
\be
{\boldsymbol\omega}_\mu(\phi, \cL_\xi \phi) = - 2 \pi \, \partial_{\tau_0} \, j_\mu \,.
\ee
This can be checked explicitly from the formulas given above. It can also been argued from the fact that $\partial_{\tau_{0}}$ acts as $i \partial_t$ on $\phi$ while it acts as $-i \partial_t$ on $\phi^\dag$ in (\ref{eq:sol}), due to the general form of the bulk-to-boundary propagator. Indeed a simple calculation shows that
\be
\label{eq:omega=j}
\boldsymbol\omega_\mu(\phi, \cL_\xi\phi) = \Bigl( (\partial_\mu \phi) \, \bigl( 2\pi\partial_t\phi^\dag \bigr) - \bigl( 2\pi \partial_\mu \partial_t\phi \bigr) \, \phi^\dag \Bigr) + \text{h.c.} = - 2\pi \, \partial_{\tau_0} \, j_\mu \,.
\ee
This equation is reminiscent of the general analysis carried out in the context of black hole mechanics and Noether charges associated with the symmetries on the black-hole horizon \cite{Wald:1993nt, Iyer:1994ys}. It would be interesting to investigate this point more systematically.

Notice that $\tau_0$ parametrizes a one-parameter family of solutions as in \eqref{eq:sol}, with delta-function sources. One could generalize this setting to a one-parameter family of smeared complex sources labelled by a parameter $\alpha$:
\be
\label{eq:generalsource}
\lambda_\alpha(\tau, Y) = \lambda_0( \tau-\alpha, Y) \,, \qquad\qquad \lambda^*_\alpha(\tau, Y) = \lambda^*_0(\tau+\alpha, Y) \,,
\ee
where $\lambda_0$ and $\lambda_0^*$ are some reference sources. The effect of $\alpha$ is to shift the boundary conditions along the modular flow. The bulk solutions are then labelled by $\alpha$ as follows:
\be
\phi_\alpha(\rho, t, Y) = \int\! d\tau' \, dY' \; K_E \bigl( \rho, it, Y \bigm| \tau', Y' \bigr) \, \lambda_\alpha(\tau', Y') \,.
\ee
Since $\partial_\tau \lambda_\alpha = - \partial_\alpha \lambda_\alpha$ and $K_E$ has translational invariance, one can trade the derivatives with respect to $t$, $\tau'$ and $\alpha$ using integration by parts:
\be
\cL_\xi \, \phi_\alpha = - 2 \pi i \, \partial_\alpha \phi_\alpha \;,\qquad\qquad \cL_\xi \, \phi^\dag_\alpha = 2 \pi i \, \partial_\alpha \phi^\dag_\alpha \,.
\ee
With this, the holographic answer can be written as
\be
\Delta S^{(2)} = - 2\pi \, \partial_\alpha \int_\Sigma \star \, j_\alpha \,,
\ee
where both sides are of order $\lambda^2$. In other words, the holographic formula (\ref{final formula from holography}) for the quadratic entanglement asymmetry can be interpreted as the derivative of the bulk charge of the entanglement wedge, with respect to a modular flow of the sources that create the state. It would be interesting to better understand the physical meaning of this observation.



\section*{Acknowledgments}

We are grateful to Filiberto Ares, Pasquale Calabrese, Laurent Freidel,  Giovanni Galati, Max Metlitski and Sridip Pal for helpful discussions.
We acknowledge support
by the ERC-COG grant NP-QFT No.~864583 ``Non-perturbative dynamics of quantum fields: from new deconfined phases of matter to quantum black holes'',
by the MUR-FARE grant EmGrav No.~R20E8NR3HX ``The Emergence of Quantum Gravity from Strong Coupling Dynamics'',
by the MUR-PRIN2022 grant No.~2022NY2MXY, and
by the INFN ``Iniziativa Specifica ST\&FI''.

\appendix


\section{Renyi asymmetries in 2d CFTs}
\label{app: asymmetry from replicas}

We compute the Renyi asymmetries in (\ref{Renyi asymmetries 2d CFT}) for a few semi-integer values of $\Delta$. In particular we have to evaluate the sums
\be
\label{summation Pn}
P_n = \sum_{j\neq k}^n \frac1{ \bigl\lvert 2 \sin \bigl[ \frac{\pi}n \bigl( j - k + \frac{\tau_0}\pi \bigr) \bigr] \bigr\rvert^{2\Delta} } \,.
\ee
We use the Mellin transform (\ref{Mellin transform}) of $1/\!\sin(\pi x)$.

\paragraph{Case \tps{\matht{\Delta = 1/2}}{Delta = 1/2}.} Taking into account that $\tau_0 \in (0, \pi)$ and therefore for $j>k$ the argument of the sine is between 0 and $\pi$ while for $j<k$ it is between $-\pi$ and 0, we have
\bea
\Delta S^{(2)}_n &= \frac{n^{-1}}{n-1} \, \frac1r \int\! \frac{ds}{2\pi \, (1+e^s)} \left[ \, \sum_{j>k}^n e^{\frac sn \left( j - k + \frac{\tau_0}\pi \right)} + \sum_{j<k}^n e^{\frac sn \left( j - k + \frac{\tau_0}\pi + n \right)} \right] \\
&= \frac{n^{-1}}{n-1} \, \frac1r \int\! ds \, \frac{ n \, (e^{s/n} - e^s) }{ 2\pi \, (1+e^s) (1-e^{s/n}) } \, e^{\frac{s \tau_0}{n \pi}} \,.
\eea
We can now take the limit $\Delta S^{(2)} = \lim_{n \to1} \Delta S^{(2)}_n$ yielding
\be
\Delta S^{(2)} = \frac1r \int\! \frac{ds \, s \, e^{s\tau_0/\pi} }{ 4\pi \sinh(s)} = \frac1r \, \frac{\pi}{8\cos^2 \bigl( \frac{\tau_0}2 \bigr)} \,.
\ee

\paragraph{Case \tps{\matht{\Delta =1}}{Delta = 1}.} We proceed as before, however this time there are two integrations:
\bea
P_n &= \frac{1}{4\pi^2} \!\int\! \frac{ ds_1 \, ds_2 }{ (1+e^{s_1}) (1 + e^{s_2})} \, \left[ \, \sum_{j>k}^n e^{ (s_1+s_2) \left( j-k + \frac{\tau_0}\pi \right)/n } + \sum_{j<k}^n e^{ (s_1+s_2) \left( j-k + \frac{\tau_0}\pi + n \right)/n}  \right] \\
&= \frac{1}{4\pi^2} \!\int\! ds_1 \, ds_2 \; \frac{ n \, \bigl( e^{ (s_1+s_2)/n } - e^{s_1+s_2} \bigr) }{ (1+e^{s_1})(1+e^{s_2})(1-e^{ (s_1+s_2)/n}) } \, e^{ \frac{ (s_1+s_2) \tau_0}{ n \pi }} \,.
\eea
The integral can be computed analytically. First we change variables setting $s_2 = s-s_1$ and perform the integral in $s_1$, obtaining
\be
P_n = \int\! ds \, \frac{ns}{8\pi^2} \, e^{\frac{ s\tau_0}{n\pi}} \, \biggl( \frac1{\tanh(s/2n)} - \frac1{\tanh(s/2)} \biggr) \,.
\ee
Then we shift the integration contour from $\bR$ to $\bR + \pi i$ by redefining $s \to s + \pi i$, as this does not cross any pole of the integrand. We perform indefinite integration in $\tau_0$ to simplify the expression, and finally integrate in $s$. We obtain
\be
P_n = \partial_{\tau_0} \biggl[ \frac{n^2}4 \biggl( \frac1{\tan(\tau_0/n)} - \frac{n}{\tan(\tau_0)} - i (n-1) \biggr) \biggr] = \frac{n^3}{4 \sin^2(\tau_0)} - \frac{n}{4\sin^2(\tau_0/n)} \,.
\ee
This yields the Renyi asymmetry
\be
\Delta S^{(2)}_n = \frac1{4r^2} \, \frac{n}{n-1} \biggl( \frac1{\sin^2(\tau_0)} - \frac1{n^2 \sin^2(\tau_0/n)} \biggr)
\ee
and the entanglement asymmetry
\be
\Delta S^{(2)} = \frac1{2r^2 \sin^2(\tau_0)} \biggl( 1 - \frac{\tau_0}{\tan(\tau_0)} \biggr) \,.
\ee

\paragraph{Case \tps{\matht{\Delta = 3/2}}{Delta = 3/2}.} This time the Renyi asymmetry is given by three integrations:
\be
\Delta S^{(2)}_n = \frac{n^{-2}}{n-1} \, \frac1{r^3} \int\! \frac{ ds_1 \, ds_2 \, ds_3 }{ 8\pi^3 \, (1+e^{s_1}) (1+e^{s_2}) (1+e^{s_3}) } \, \frac{ (e^{s/n} - e^s) }{ (1 - e^{s/n})} \, e^{\frac{ s\tau_0}{n\pi}} \,,
\ee
where $s = s_1 + s_2 + s_3$. Taking the limit $n\to1$ we obtain the entanglement asymmetry
\be
\Delta S^{(2)} = \frac1{8\pi^3 r^3} \int\! \frac{ ds_1 \, ds_2 \, ds_3 }{ (1+e^{s_1}) (1+e^{s_2}) (1+e^{s_3}) } \, \frac{s \, e^{s \tau_0/\pi} }{ (1 - e^{-s}) } \,.
\ee
It is convenient to change variables to $s_1, s_2, s$.
The integral in $s_1$ can be easily performed:
\be
\Delta S^{(2)} = \frac1{8\pi^3 r^3} \int\! ds_2 \, ds \, \frac{ s \, (s-s_2) \, e^{s\tau_0/\pi} }{ (1 + e^{-s_2}) (e^s - e^{s_2}) (1-e^{-s}) } \,.
\ee
Then we shift the integration contour $s \to s + \pi i$. The integral in $s_2$ gives
\be
\Delta S^{(2)} = \frac{e^{i\tau_0}}{ 8\pi^3 r^3} \int\! ds \, \frac{ s (s+\pi i) (s+2\pi i) \, e^{s\tau_0/\pi} }{ 2 \, (1+e^{-s}) (1 - e^s) } \,.
\ee
The final integral gives  the entanglement asymmetry
\be
\Delta S^{(2)} = \frac{3\pi}{128 \, r^3 \cos^4 \bigl( \frac{\tau_0}2 \bigr) } \,.
\ee

\paragraph{Case \tps{\matht{\Delta = 2}}{Delta = 2}.} The Renyi asymmetry is given by four integrations. After taking the limit $n\to1$ and changing variables to $s = s_1 + s_2 + s_3 + s_4$, we obtain the expression
\be
\Delta S^{(2)} = \frac1{(2\pi r)^4} \int\! \frac{ ds\, ds_1 \, ds_2 \, ds_3 \quad s \, e^{s_1 + s_2 + s_3} \, e^{s \tau_0 / \pi} }{ (1+e^{s_1}) (1+e^{s_2}) (1+ e^{s_3}) ( e^{s_1 + s_2 + s_3} + e^s) (1 - e^{-s}) } \,.
\ee
We perform the integral in $s_1$, and then shift the integration contour $s \to s + \pi i$:
\be
\Delta S^{(2)} = \frac{ e^{i\tau_0}}{(2\pi r)^4} \int\! ds\, ds_2 \, ds_3 \, \frac{ (s+\pi i) (s_2 + s_3 - s - \pi i) \, e^{s\tau_0/\pi} }{ (1+e^{-s_2}) (1+ e^{-s_3}) ( e^{s_2 + s_3} + e^s ) (1 + e^{-s}) } \,.
\ee
We perform the integral in $s_2$, and then shift the integration contour back as $s \to s - \pi i$:
\be
\Delta S^{(2)} = \frac1{(2\pi r)^4} \int\! ds\, ds_3 \, \frac{ s (s-s_3 - \pi i)(s - s_3 + \pi i) \, e^{s \tau_0/\pi} }{ 2 (1+e^{-s_3}) (e^{s_3} + e^s) (1 - e^{-s}) } \,.
\ee
We perform the integral in $s_3$:
\be
\Delta S^{(2)} = \frac1{(2\pi r)^4} \int\! ds \, \frac{s^2 (s^2 + 4\pi^2) \, e^{s\tau_0/\pi} }{ 6 (e^s -1) (1 - e^{-s}) } \,.
\ee
The final integration gives the entanglement asymmetry
\be
\Delta S^{(2)} = \frac1{4r^4 \sin^4(\tau_0)} \biggl( 1 - \frac13 \sin^2(\tau_0) - \frac{\tau_0}{ \tan(\tau_0)} \biggr) \,.
\ee


\section{Computation from relative entropy}
\label{app:rindlerasymmetry}

Let us give some details on the evaluation of the integral (\ref{asymmetry integral}):
\be
\label{RelativeEntropyIntegral}
I_\Delta = - \int_\bR ds \; \frac{1}{ 4 \,\sinh^2\bigl( \frac s2 - i \varepsilon) \, \Bigl[ - 4 \sinh^2 \bigl( \frac s2 - i \tau_0 \bigr) \Bigr]^\Delta} \,,
\ee
in terms of the function $I_\Delta(x)$ in (\ref{def function I}) with $x = \tan(\tau_0/2)$.

First notice that, calling $f(s)$ the integrand in (\ref{RelativeEntropyIntegral}), it satisfies $f(-s^*) = f(s)^*$. Since the operation $s \to -s^*$ maps the part of the contour with $s>0$ to the part with $s<0$, this implies that the integral can be written as
\be
I_\Delta = \int_0^\infty \re\bigl[ f(s) \bigr] \,,
\ee
showing that the integral is always real. The integrand has poles at $s= 2 i \varepsilon + 2 \pi i k$ and $s = 2 i \tau_0 + 2 \pi i k$ for any $k \in \bZ$. We thus shift the contour as $s \to s + 2i \tau_0 - \pi i$. For $0 < \tau_0 < \frac\pi2$ the contour is moved downward and it does not cross any pole of $f(s)$. On the contrary, for $\frac\pi2 < \tau_0 < \pi$ the contour is moved upward and it crosses the pole at $s = 2 i \varepsilon$ where we pick up a residue. We thus obtain the expression
\be
\label{formula I_Delta with residue}
I_\Delta = J_\Delta - \delta_{\frac\pi2 < \tau_0 < \pi} \; \frac{ 2 \pi \, \Delta}{ 4^\Delta \sin^{2\Delta}(\tau_0) \tan(\tau_0) }
\ee
where
\be
J_\Delta = \int_0^\infty \! \re \Biggl[ \frac{ 2^{-2\Delta -1} \; ds }{ \cosh^2\bigl( \frac s2 + i \tau_0 \bigr) \cosh^{2\Delta} \bigl( \frac s2 \bigr) } \Biggr]
= \int_0^\infty \! \frac{ \bigl( 1 + \cos(2\tau_0) \cosh(s) \bigr) \; ds }{ 2^\Delta \bigl( \cos(2\tau_0) + \cosh(s) \bigr)^2 \bigl( 1 + \cosh(s) \bigr)^\Delta } \,.
\ee
To compute the integral we perform the change of variable $t = \tanh^2\bigl( \frac s2 \bigr)$:
\be
J_\Delta = \int_0^1 dt \; \frac{ (1+X) (1-t X) (1-t)^\Delta }{ 2^{2\Delta + 1} \, \sqrt{t} \; (1+t X)^2 }
\ee
where we defined the parameter $X = \tan^2(\tau_0)$. This can be integrated as a hypergeometric function
\be
J_\Delta = \frac{ \sqrt{\pi} \, \Gamma(\Delta + 1) }{ 4^\Delta \, \Gamma \bigl( \Delta + \frac12 \bigr) } \, \biggl( 1 - \frac{\Delta}{\Delta + \frac12} \, {}_2F_1 \Bigl[ \tfrac12, 1, \Delta + \tfrac32; -X \Bigr] \biggr) \,.
\ee
The hypergeometric function has a branch cut from 1 to infinity, conventionally taken along the positive real axis. When $0 < \tau_0 < \frac\pi2$ the argument $-X$ of the hypergeometric function goes from 0 to infinity along the negative real axis. For $\tau_0 = \frac\pi2$ the argument reaches the branch point at infinity and it moves to the second sheet. When $\frac\pi2 < \tau_0 < \pi$ the argument goes from infinity to zero on the second sheet along the negative real axis. The fact that the argument is on the second sheet is implemented by the second term in (\ref{formula I_Delta with residue}).

In order to obtain an equivalent expression that does not require to move between different sheets, it is useful to use a different parameter $x = \tan\bigl( \frac{\tau_0}2 \bigr)$, related to $X$ by
\be
\label{transfXtox}
X = \frac{ 4x^2 }{ (x^2-1)^2} \,.
\ee
Then one can use hypergeometric transformation identities (see for example \cite{andrewsSpecialFunctions1999}) to rewrite the answer in terms of $x$, obtaining (\ref{def function I}). A simple way to implement the transformation is to use the fact that the hypergeometric function $y(z) = {}_2 F_1(a,b,c;z)$ satisfies a second-order differential equation:
\be
z(1-z) \, y''(z) + \bigl( c - (a+b+1)z \bigr) \, y'(z) - ab \, y(z) = 0 \,.
\ee
This implies that the function $I_\Delta \equiv F(X)$ satisfies the differential equation
\be
2 X(X+1) \, F''(X) + (2\Delta +5 X +3) \, F'(X) + F(X)  = \frac{ \sqrt{\pi} \, \Gamma(\Delta + 1) }{4^\Delta \Gamma \bigl( \Delta + \frac12 \bigr) } \,.
\ee
We then use the change of variable \eqref{transfXtox} to write a differential equation for the function $I_\Delta \equiv f(x)$ in terms of the parameter $x$:
\be
\frac{(x^2-1)^2}8 \, f''(x) - \frac{(x^2-1) \bigl( 1 + 3x^2 + (x^2-1)^2 \Delta) }{ 4x(x^2+1)} \, f'(x) + f(x) = \frac{ \sqrt{\pi} \, \Gamma(\Delta + 1) }{4^\Delta \Gamma \bigl( \Delta + \frac12 \bigr) } \,.
\ee
Solving this equation again gives the expression in \eqref{def function I}.


\section{Asymmetry from twist operators}
\label{app: asymmetry from twist}

As in \cite{Calabrese:2004eu}, the Renyi asymmetries can be computed by making use of replicated theories, as opposed to replicated geometries. Given a theory $\cS$, we consider a theory $\cS^{(n)}$ made of $n$ decoupled copies of $\cS$. Since $\cS^{(n)}$ has a permutation symmetry that exchanges the copies, it also has a codimension-two twist operator $\cT_n$ such that when we circle around it we go from one copy to the next, cyclicly. More precisely, $\cT_n$ is the (non-topological) boundary of a codimension-one topological symmetry defect such that the fields on the two sides are glued with a cyclic permutation. If theory $\cS$ has a $U(1)$ symmetry, the previous symmetry defect can be combined with other symmetry defects that implement independent $U(1)$ actions on each copy of $\cS$ in $\cS^{(n)}$. This generates a dressed twist operator $\cT_{n,\gamma}$ that permutes the copies and besides acts with $e^{i \gamma_j Q_A}$ on the $j$-th copy. The twist operators relevant for the computation of Renyi asymmetries have $\sum_j \gamma_j = 0 \text{ mod } 2\pi$. Similarly to (\ref{Renyi asymmetry QFT}), the Renyi asymmetry is then given by a ratio of correlation functions, one with the insertion of the dressed twist fields, and one with the standard twist fields:
\be
\label{Renyi asymmetries twist ops}
\Delta S_n = \frac1{1-n} \log \frac{ \frac1{(2\pi)^{n-1}} \int\! d\vec\gamma \; \bigl\langle \cT_{n,\gamma}[\partial A] \,\ldots\, \bigr\rangle }{ \bigl\langle \cT_n[\partial A] \,\ldots\, \bigr\rangle } \,.
\ee
The twist fields are placed along the boundaries of the cut $A$. The dots stand for possible operator insertions used to construct the state.

In two dimensions the symmetry defects are lines and the twist operators are pointlike, therefore (\ref{Renyi asymmetries twist ops}) boils down to the computation of certain correlation functions. In this appendix we use (\ref{Renyi asymmetries twist ops}) to compute the Renyi asymmetry of a 2d free Dirac fermion $\Psi$ with respect to its vector-like $U(1)$ symmetry, in coherent states generated by the primary%
\footnote{It can be written as $V = \Psi^\sT C \Psi$ where $C$ is the charge conjugation matrix.}
$V = \psi_L \psi_R$ with $\Delta =1$, for an interval $A = [-\ell, \ell]$. This computation will reproduce (\ref{asymmetry interval}) for $\Delta=1$, as well as a conformal transformation of (\ref{Renyi asymm Delta = 1}). 

The theory $\cS^{(n)}$ is given by $n$ copies $\Psi_k$ ($k=1, \dots, n$) of the 2d free Dirac fermion. The twist operator $\cT_{n,\gamma}(u)$ implements the boundary condition%
\footnote{This computation is adapted from Section~2.4 of \cite{Belin:2013uta}. The minus signs in the second line are due to the fact that we are dealing with fermions, as in thermal partition functions, as explained in Section~2 of \cite{Casini:2005rm}.}
\bea
\Psi_k \bigl( u + e^{2\pi i} w \bigr) &= e^{i\gamma_k} \, \Psi_{k+1}( u + w) &&\text{for } k = 1, \dots, n-1 \\
\Psi_n \bigl( u + e^{2\pi i} w \bigr) &= (-1)^{n+1} e^{-i (\gamma_1 + \ldots + \gamma_{n-1})} \Psi_1( u + w)
\eea
around $u$. The twist operator $\cT_{n,\gamma}^{\,\dag}$ implements the opposite boundary condition. Writing $\gamma_k = \alpha_{k+1} - \alpha_k$, the boundary conditions are diagonalized by the fields
\be
\label{diagonalizing basis}
\tilde\Psi_l = \frac1{\sqrt{n}} \sum_{k=1}^n \exp\biggl( - 2\pi i \, \frac{2l - n - 1}{2n} \, k + i \, \alpha_k \biggr) \, \Psi_k
\ee
in the sense that
\be
\tilde\Psi_l \bigl( u + e^{2\pi i} w \big) = e^{2\pi i m_l} \, \tilde\Psi_l(u + w) \qquad\text{with}\qquad m_l = \frac{2l - n-1}{2n} \,.
\ee
Notice that in the diagonalizing basis the twists do not depend on the $\gamma_k$'s.
We decompose the Dirac fermion $\Psi = \smat{ \psi_L \\ \psi_R}$ into left- and right-moving parts, and the two fields $\psi_L$ and $\psi_R$ have the same charge under $U(1)$. Then the scalar primary we use to construct the excited coherent states is $V = \psi_L \psi_R$. 

In order to construct the twist operators we use bosonization. For each copy $\tilde\Psi_l$ we introduce a compact scalar%
\footnote{Take $S = \frac1{8\pi} \!\int\! d^2\sigma \, \partial_\mu \varphi \partial^\mu \varphi$ with $\varphi \cong \varphi + 2\pi R$. The operators $\cO_{n,w} = e^{in \varphi /R} e^{iw R \tilde\varphi/2}$ (where the dual scalar $\tilde\varphi$ is defined by $\partial_\mu \tilde\varphi = \epsilon_{\mu\nu} \partial^\nu\varphi$) have conformal dimensions and spin
\be
h_{n,w} = \frac12 \biggl( \frac nR + \frac{wR}2 \biggr)^2 \;,\qquad \bar h_{n,w} = \frac12 \biggl( \frac nR - \frac{wR}2 \biggr)^2 \,\qquad \Delta_{n,w} = \frac{n^2}{R^2} + \frac{w^2R^2}4 \;,\qquad s_{n,w} = nw \;.
\ee
T-duality maps $R \leftrightarrow 2/R$. The case $R=1$ gives the bosonization of the Dirac fermion. Indeed $\cO_{\frac12,1} = \tilde\psi_L$ and $\cO_{-\frac12, -1} = \tilde\psi_L^\dag$ are the left-moving fermion while $\cO_{\frac12, -1} = \tilde\psi_R$ and $\cO_{-\frac12,1} = \tilde\psi_R^\dag$ are the right-moving fermion.}
$\varphi \cong \varphi + 2\pi$ and decompose it into left- and right-moving parts: $\varphi = \varphi_L + \varphi_R$ and $\tilde\varphi = \varphi_L - \varphi_R$. An operator $\cO = e^{i \epsilon_L \varphi_L} e^{i \epsilon_R \varphi_R}$ has dimensions $(h, \bar h) = \bigl( \epsilon_L^2/2, \epsilon_R^2/2 \bigr)$. It follows that the fermions are
\be
\tilde\psi_L = e^{i \varphi_L} \;,\qquad \tilde\psi_L^\dag = e^{-i \varphi_L} \;,\qquad \tilde\psi_R = e^{i\varphi_R} \;,\qquad \tilde\psi_R^\dag = e^{-i \varphi_R} \;.
\ee
The current that implements the shift symmetry $\varphi \to \varphi + \text{const}$ is $J_\mu = \frac1{2\pi} \partial_\mu \varphi$ and the charge operator is $Q = \int\! \star J = \frac1{2\pi} \!\int\! d\tilde\varphi$. The twist operator lives at the end of a symmetry defect. Thus the operator that implements a twist by $e^{2\pi i \epsilon}$ along the positive real axis, which lives at the left end of the line $e^{2\pi i \epsilon Q}$, is $e^{i \epsilon\tilde\varphi} = e^{i\epsilon \varphi_L} e^{-i \epsilon \varphi_R}$ with dimensions $(h, \bar h) = \bigl( \frac{\epsilon^2}2, \frac{\epsilon^2}2 \bigr)$ and $\Delta = \epsilon^2$. To resolve possible ambiguities, we insist that the twist operator is the ground state of the twisted sector. This means that we take $\epsilon \in \bigl[ -\frac12, \frac12 \bigr]$. Let us call $\sigma_l$ the twist operator in the $l$-th copy that implements a twist by $e^{2\pi i m_l}$. The full twist operator is
\be
\label{expansion twist operators}
\cT_{n,\gamma} = \prod\nolimits_{l=1}^n \sigma_l = \prod\nolimits_{l=1}^n e^{i m_l \tilde\varphi_l}
\ee
and its conformal dimension is
\be
\Delta\bigl[ \cT_{n,\gamma} \bigr] = \sum\nolimits_{l=1}^n m_l^2 = \frac{n^2-1}{12n} \,.
\ee
This reproduces the dimension of the twist fields in the case of central charge $c=1$ \cite{Calabrese:2004eu} and shows that such a dimension does not depend on the parameters $\gamma_k$ \cite{Capizzi:2023yka}.

Let us now compute the Renyi asymmetries of the coherent states created by $V$, for a finite subset $A = [-\ell, \ell]$ as in Section~\ref{sec: finite subregion}. For simplicity we take the insertions along the imaginary axis: $w_- = -i\eta$ and $w_+ = i\eta$. The correlator we need to compute is thus
\be
\cA = \biggl\langle \cT_{n,\gamma}(-\ell) \; \cT_{n, \gamma}^{\,\dag}(\ell) \; \biggl(\, \prod_{j=1}^n e^{i \lambda V_j(-i\eta)} \biggr) \; \biggl(\, \prod_{j=1}^n e^{-i \lambda V^\dag_j (i\eta)} \biggr) \biggr\rangle_{\cS^{(n)}, \bC} \,.
\ee
Here $V_j$ is the primary in the $j$-th copy, and we put identical insertions in each copy. In (\ref{Renyi asymmetries twist ops}) there is one such correlator in the numerator integrated over $\vec\gamma$, and one such correlator in the denominator with $\vec\gamma =0$.
We expand the correlator in $\lambda$. At zero-th order we have
\be
\cA^{(0)} = \bigl\langle \cT_{n,\gamma}(-\ell) \; \cT_{n,\gamma}^{\,\dag}(\ell) \bigr\rangle = \frac1{(2\ell)^{2\Delta_\cT}}
\ee
where $\Delta_\cT = (n^2-1)/12n$. At second order we have
\be
\label{correlator A2}
\cA^{(2)} = \Bigl\langle \cT_{n,\gamma}^{\phantom\dag} \, \cT_{n,\gamma}^{\,\dag} \sum\nolimits_{k_1, k_2} \Bigl( V_{k_1}^{\phantom\dag} \, V^\dag_{k_2} \Bigr) \Bigr\rangle \;,
\ee
where we left the dependence on positions implicit in order not to clutter. We expand each of the fields in the diagonalizing basis:%
\footnote{The inverse to (\ref{diagonalizing basis}) is $\ds \Psi_k = \frac1{\sqrt{n}} \sum_{l=1}^n \exp\biggl( 2\pi i \, \frac{2l-n-1}{2n}\, k - i \, \alpha_k \biggr) \, \tilde\Psi_l$.}
the twist operators are as in (\ref{expansion twist operators}) while
\bea
V_{k_1} &= \frac1n \, \sum_{j_1, j_2} \exp\biggl( \phantom{+} \frac{2\pi i}n \bigl( j_1 + j_2 - n- 1 \bigr) k_1 - 2 i \alpha_{k_1} \biggr) \, \tilde\psi_{L j_1} \tilde\psi_{R j_2} \\
V^\dag_{k_2} &= \frac1n \, \sum_{j_3, j_4} \exp\biggl( - \frac{2\pi i}n \bigl( j_3 + j_4 - n- 1 \bigr) k_2 + 2i \alpha_{k_2} \biggr) \, \tilde\psi^\dag_{L j_3} \tilde \psi^\dag_{R j_4} \,.
\eea
When we substitute in the correlator (\ref{correlator A2}) we obtain six sums. However if we insert $\tilde \psi_{L j_1}$ then there must also be $\tilde \psi^\dag_{L j_1}$ or else the correlator vanishes, which means that we restrict to $j_3 = j_1$. Similarly we restrict to $j_4 = j_2$. We thus find
\bea
\cA^{(2)} &= \frac1{n^2} \sum_{k_1, k_2, j_1, j_2} \exp\biggl[ \frac{2\pi i}n \bigl( j_1 + j_2 - n - 1 \bigr)(k_1 -k_2) - 2i \bigl( \alpha_{k_1} - \alpha_{k_2} \bigr) \biggr] \times {} \\
&\quad \times \biggl\langle \biggl( \prod\nolimits_l \, e^{i m_l \tilde\varphi_l} \, e^{-i m_l \tilde\varphi_l} \biggr) \, \tilde\psi_{L j_1} \tilde\psi_{R j_2} \tilde\psi^\dag_{L j_1} \tilde\psi^\dag_{R j_2} \biggr\rangle \;.
\eea
In the numerator the integrals over $\vec\gamma$, equivalent to integrals over $\vec\alpha$, impose $k_1 = k_2$. In the denominator there are no such integrals. So at order $\lambda^2$ we are left with a sum over $k_1 \neq k_2$ (and no $\alpha$'s). We need to compute each term of this sum.

To compute the correlators we use the Coulomb gas formalism, \ie, we separately compute the left- and right-moving contributions. The holomorphic part of a correlator is
\be
\Bigl\langle e^{i \epsilon_1 \varphi_L} (z_1) \dots e^{i \epsilon_n \varphi_L}(z_n) \Bigr\rangle = \prod\nolimits_{i<j}^n (z_i - z_j)^{\epsilon_i \epsilon_j} \;.
\ee
The left-moving part of the correlator is given by
\begin{align}
\cA^{(2)}_{L, k_1k_2} &= \frac1n \sum_{j_1=1}^n \biggl\langle \biggl( \prod\nolimits_l \, e^{i m_l \varphi_{Ll}}(-\ell) \, e^{-i m_l \varphi_{Ll}}(\ell) \biggr) \, e^{i \varphi_{L j_1}}(-i\eta) \, e^{-i \varphi_{L j_1}}(i\eta) \biggr\rangle \; e^{2\pi i \, \frac{2j_1-n-1}{2n} (k_1 - k_2) } \nn\\
&= \biggl( \, \prod_{l \neq j_1} (2\ell)^{- m_l^2} \biggr) \frac1n \sum_{j_1=1}^n \frac{(2\ell)^{- m_{j_1}^2}}{ 2i\eta } \biggl[ \frac{ ( i\eta -\ell) (- i\eta + \ell) }{ (i \eta + \ell) ( - i \eta -\ell ) } \biggr]^{m_{j_1}} e^{2\pi i \, \frac{2j_1-n-1}{2n} (k_1 - k_2) } \\
&= \frac{n^{-1}}{ (2\ell)^{\Delta_\cT} \, (2i\eta) } \sum_{j_1=1}^n e^{ 2\pi i \, \frac{2j_1 - n-1}{2n} \, \left( k_1 - k_2 + \frac{\tau_0}{\pi} \right)}
= \frac{1}{ (2\ell)^{\Delta_\cT} \, (2i\eta) } \, \frac{ \sin \bigl( \pi(k_1 - k_2) + \tau_0 \bigr) }{ n \sin\bigl( \frac{ \pi (k_1-k_2) + \tau_0 }{n} \bigr) } \;, \nn
\end{align}
where $\tau_0 = 2 \arctan(\ell/\eta)$. To correctly evaluate the second line we placed the cut of each twist operator on the right, and this unambiguously fixes all phases. In particular:
\bea
& \bigl\langle e^{i \epsilon \varphi_L}(-\ell) \, e^{-i\epsilon \varphi_L}(\ell) \, e^{i\varphi_L}(-i\eta_1) \, e^{-i\varphi_L}(i\eta_2) \bigr\rangle = \frac1{(2\ell)^{\epsilon^2} \, i (\eta_1 + \eta_2)} \, \frac{ (i\eta_2 - \ell)^\epsilon (-i\eta_1 + \ell)^\epsilon }{ (i\eta_2 + \ell)^\epsilon (-i\eta_1 - \ell)^\epsilon} \\
&\qquad = \frac1{(2\ell)^{\epsilon^2} \, i (\eta_1 + \eta_2)} \, \frac{ e^{i(\pi - \theta_2)\epsilon} \, e^{i (2\pi - \theta_1) \epsilon} }{ e^{i \theta_2 \epsilon} \;\; e^{i(\pi + \theta_1)} } = \frac1{(2\ell)^{\epsilon^2} \, i (\eta_1 + \eta_2)} \, e^{i (2\pi - 2\theta_1 - 2\theta_2) \epsilon} \;,
\eea
where $\tan \theta_j = \eta_j/\ell$ and $0 < \theta_j < \frac\pi2$. Here it is important and all angles are taken from the cut (towards the right) going counterclockwise. See Fig.~\ref{fig: 4-point function}. In our example we take $\theta_1 = \theta_2 \equiv \theta$ and, defining $\tan(\tau_0/2) = \ell/\eta$, we have $\theta = (\pi- \tau_0)/2$.
\begin{figure}[t]
\centering
\begin{tikzpicture}
	\draw [gray!50!white] (-2, -0.04) -- (0, 2) -- (2, 0.04) -- (0, -2) -- cycle;
	\draw [thick, red!80!black] (2, 0.04) -- (4, 0.04);
	\filldraw (2, 0.04) circle [radius = 0.04] node [shift={(-80: 0.4)}] {\small$\ell$};
	\draw [thick, red!80!black] (-2, -0.04) -- (4, -0.04);
	\filldraw (-2, -0.04) circle [radius = 0.04] node [shift={(-120: 0.4)}] {\small$-\ell$};
	\filldraw (0, 2) circle [radius = 0.04] node [xshift = 15] {\small$i\eta_2$};
	\filldraw (0, -2) circle [radius = 0.04] node [xshift = 18] {\small$-i\eta_1$};
	\draw [->] (-2, -0.04) +(5: 1) arc [radius = 1, start angle = 5, end angle = 42] node [pos = 0.5, shift={(10: 0.25)}] {\small$\theta_2$};
	\draw [->] (-2, -0.04) +(5: 0.8) arc [radius = 0.8, start angle = 5, end angle = 312] node [pos = 0.85, shift={(-90: 0.25)}] {\small$2\pi - \theta_1$};
	\draw [->] (2, 0.04) +(5: 1) arc [radius = 1, start angle = 5, end angle = 132] node [pos = 0.4, shift={(10: 0.7)}] {\small$\pi - \theta_2$};
	\draw [->] (2, 0.04) +(5: 0.8) arc [radius = 0.8, start angle = 5, end angle = 222] node [pos = 0.95, shift={(180: 0.6)}] {\small$\pi + \theta_1$};
\end{tikzpicture}
\caption{\label{fig: 4-point function}%
Phases involved in the 4-point function $\bigl\langle e^{i \epsilon \varphi_L}(-\ell) \, e^{-i\epsilon \varphi_L}(\ell) \, e^{i\varphi_L}(-i\eta_1) \, e^{-i\varphi_L}(i\eta_2) \bigr\rangle$.}
\end{figure}
The right-moving part of the correlator is similar, however: $\varphi_{Lj} \to \varphi_{Rj}$; the twist operator is $e^{-i\epsilon\varphi_R}$; all dependences are on the anti-holomorphic coordinates. This is the same as mapping $\epsilon \to - \epsilon$ and $\eta \to - \eta$, and it gives:
\be
\cA^{(2)}_{R, k_1k_2} = \frac{1}{ (2\ell)^{\Delta_\cT} \, (-2i\eta) } \, \frac{ \sin \bigl( \pi(k_1 - k_2) + \tau_0 \bigr) }{ n \sin\bigl( \frac{ \pi (k_1-k_2) + \tau_0}{n} \bigr) } \;.
\ee
Putting the two contributions together, we find
\be
\cA^{(2)}_{k_1 k_2} = \frac1{(2\ell)^{2\Delta_\cT} \, 4\eta^2} \, \frac{ \sin^2 ( \tau_0 ) }{ n^2 \sin^2 \bigl( \frac{ \pi (k_1-k_2) + \tau_0}{n} \bigr) } \;.
\ee

The expansion of (\ref{Renyi asymmetries twist ops}) in $\lambda$ gives
\bea
\Delta S_n
&= \frac1{1-n} \log \frac{ \cA^{(0)} + \lambda^2 \frac1{(2\pi)^n} \!\int\! d\vec\alpha \, \sum_{k_1 k_2} \cA^{(2)}_{k_1 k_2}(\vec \alpha) + O(\lambda^4) }{ \cA^{(0)} + \lambda^2 \sum_{k_1k_2} \cA^{(2)}_{k_1 k_2}(0) + O(\lambda^4) } \\
&= \frac1{1-n} \log \frac{ 1 + \frac{\lambda^2}{\cA^{(0)}} \sum_k \cA^{(2)}_{kk} + O(\lambda^4) }{ 1 + \frac{\lambda^2}{\cA^{(0)}} \sum_{k_1k_2} \cA^{(2)}_{k_1 k_2} + O(\lambda^4) } = \frac{ \lambda^2}{n-1} \sum_{k_1 \neq k_2}^n \frac{\cA^{(2)}_{k_1 k_2}}{ \cA^{(0)} } + O(\lambda^4) \;.
\eea
The sum over $k_1 \neq k_2$ is essentially the same one that we already encountered in (\ref{summation Pn}) for $\Delta = 1$. Therefore at quadratic order:
\be
\Delta S^{(2)}_n = \frac1{n-1} \, \frac1{4\eta^2} \biggl( n - \frac{ \sin^2(\tau_0) }{ n \sin^2(\tau_0/n) } \biggr) \;.
\ee
This is indeed the Renyi asymmetry of an interval in the case $\Delta = 1$, that one obtains by performing a conformal transformation from the Rindler geometry and using (\ref{Renyi asymm Delta = 1}). The limit $n\to1$ gives $\Delta S^{(2)}_\text{disk} = \bigl( 1 - \tau_0 \cot(\tau_0) \bigr)/2\eta^2$ that reproduces (\ref{asymmetry interval}).

\section{Embedding formalism}
\label{app: embedding}

In Poincar\'e coordinates, the metric of Euclidean AdS$_{d+1}$ is
\be
\label{AdS in Poincare coordinates}
ds^2 = \frac{ dz^2 + dx_1^2 + \ldots + dx_d^2}{ z^2}
\ee
where $z>0$. This can be obtained from the embedding formalism by introducing constrained coordinates $P_A = (P_{-1}, P_0, \ldots, P_d)$ that satisfy
\be
P\cdot P \,\equiv\, - P_{-1}^2 + P_0^2 + \sum\nolimits_{i=1}^d P_i^2 = -1 \,,\qquad\qquad P_{-1} \geq 1 \,,
\ee
and projecting the flat metric $ds^2 = - dP_{-1}^2 + dP_0^2 + \sum_{i=1}^d dP_i^2$. This displays the $SO(d+1,1)$ isometry. The Poincar\'e coordinates are obtained from the parametrization
\be
P_{-1} = \frac{ 1 + z^2 + \vec x^{\,2}}{2z} \,,\qquad\qquad P_0 = \frac{1 - z^2 - \vec x^{\,2}}{2z} \,,\qquad\qquad P_{i=1, \ldots, d} = \frac{x_i}{z} \,.
\ee
Here $\vec x = (x_i) = (x_1, \dots, x_d)$. The conformal boundary is at $z=0$.

The Rindler coordinates appearing in (\ref{eq:rindlermetric}) are obtained from the parametrization
\be
P_{-1} = \rho\, Y_0 \,,\quad P_0 = \sqrt{\rho^2-1} \, \cos(\tau) \,,\quad P_d = \sqrt{\rho^2-1} \sin(\tau) \,,\quad P_{a=1, \ldots, d-1} = \rho\, Y_a \,,
\ee
where $\rho \geq 1$, the variable $\tau \in [0,2\pi)$ is an angle, while $(Y_0, Y_a)$ are auxiliary coordinates on $\bH^{d-1}$ that satisfy $-Y_0^2 + \sum_{a=1}^{d-1} Y_a^2 = -1$ and $Y_0 \geq 1$. We parametrize them as
\be
Y_0 = \cosh(u) \,,\qquad\qquad Y_{a=1, \ldots, d-1} = \sinh(u) \, \theta_a \,,
\ee
where $u \geq 0$ (although in the case of $d=2$ we could set $\theta_1=1$ and use $u \in \bR)$ while $\theta_a$ satisfy $\sum_{a=1}^{d-1} \theta_a^2 = 1$ and describe a sphere $S^{d-2}$ of radius 1. The induced metric is (\ref{eq:rindlermetric}):
\be
\label{AdS in Rindler coordinates}
ds^2 = \bigl( \rho^2-1 \bigr) \, d\tau^2 + \frac{d\rho^2}{\rho^2-1} + \rho^2 \, \Bigl( du^2 + \sinh^2(u) \, d\Omega_{d-2}^2 \Bigr)
\ee
where $d\Omega_{d-2}^2$ is the metric on $S^{d-2}$ induced from $ds^2 = \sum_a d\theta_a^2$. The two parametrizations of $P_A$ given above define the change of coordinates.

In Poincar\'e coordinates the bulk-to-boundary propagator is as in (\ref{bulk-to-boundary prop Poincare}):
\be
K_E \bigl( z, \vec x \bigm| \vec x^{\,\prime} \bigr) = C_\Delta \, \biggl( \frac{z}{ z^2 + (\vec x - \vec x^{\,\prime})^2} \biggr)^\Delta \,.
\ee
In order to derive the propagator in Rindler coordinates we use the identity
\bea
\label{identity embedding coord}
-2\rho\, Y \cdot Y' - 2\sqrt{\rho^2-1} \cos(\tau - \tau') &= 2 \Bigl[ P_{-1} Y_0' - P_0 \cos(\tau') - P_a Y_a' - P_d \sin(\tau') \Bigr] \\
&= \bigl( Y_0' + \cos(\tau') \bigr) \, \frac{z^2 + (\vec x - \vec x^{\,\prime})^2}z
\eea
where%
\footnote{To invert these formulas (we omit primes) one can use $\ds x_d = \frac{\sin(\tau)}{Y_0 + \cos(\tau)}$, $\ds \frac{ 1 - \vec x^{\,2}}2 = \frac{\cos(\tau)}{Y_0 + \cos(\tau)}$ from which one extracts $\ds \tan(\tau) = \frac{2x_d}{1 - \vec x^{\,2}}$ and $\ds Y_0 = \frac{ 1+\vec x^{\,2}}{ \sqrt{ (1-\vec x^{\,2})^2 + 4x_d^2 }} \geq 1$.}
\be
\label{boundary coordinate change}
x_a' = \frac{Y_a'}{Y_0' + \cos(\tau')} \,,\qquad\qquad x_d' = \frac{\sin(\tau')}{Y_0' + \cos(\tau')} \,,\qquad\qquad \vec x^{\,\prime\,2} = \frac{Y_0' - \cos(\tau')}{Y_0' + \cos(\tau')} \,.
\ee
The primed coordinates identify a point on the conformal boundary. In Poincar\'e coordinates, the metric induced on the conformal boundary at $z=0$ from (\ref{AdS in Poincare coordinates}) is the flat one of $\bR^d$, namely $ds^2_\partial = d\vec x^{\,\prime\,2}$. In these coordinates we regard $x_d'$ as the Euclidean time, and consider the spherical spatial subregion $A = \bigl\{ x'_d=0,\; \vec x^{\,\prime\,2} < 1 \bigr\}$ where the second term indicates the Euclidean norm of the spatial coordinates. The change of boundary coordinates (\ref{boundary coordinate change}) gives
\be
ds^2_\partial = d\vec x^{\,\prime\,2} = \Omega^{-2} \Bigl( d\tau^{\prime\, 2} + du^{\prime\,2} + \sinh^2(u') \, d\Omega_{d-2}^{\prime\,2} \Bigr) \,,\qquad\quad \Omega = \cosh(u') + \cos(\tau') \,.
\ee
Here $\Omega$ is precisely the conformal factor (\ref{conformal factor from holography}), also appearing in (\ref{identity embedding coord}). The metric in parenthesis is the one of $S^1 \times \bH^{d-1}$, which is also the metric on the conformal boundary at $\rho = \infty$ induced from (\ref{AdS in Rindler coordinates}). The spherical spatial subregion $A$ is mapped to a copy of $\bH^{d-1}$ at $\tau'=0$. Since the propagator transforms as a primary of dimension $\Delta$, using (\ref{identity embedding coord}) we conclude that
\be
K_E \bigl( \rho, \tau, Y \bigm| \tau', Y' \bigr) = \frac{C_\Delta}{ \Bigl( -2\rho\, Y\cdot Y' - 2 \sqrt{\rho^2-1} \cos(\tau - \tau') \Bigr)^\Delta} \,.
\ee


\bibliographystyle{ytphys}
\baselineskip=0.94\baselineskip
\bibliography{Entanglement}

\end{document}